%% file: Kanva.tex
\documentclass{article}

\usepackage{PRIMEarxiv}

\usepackage{float}
\usepackage{hyperref}
\usepackage{mdframed}
\usepackage{amssymb}
\usepackage{xspace}
\usepackage{algorithmicx}
\usepackage{algorithm}
\usepackage[noend]{algpseudocode}
\usepackage{paralist}
\usepackage{xcolor}
\usepackage{url}
\usepackage{graphicx}
\usepackage{tikz}
\usepackage{multirow}
\usepackage{siunitx}
\usepackage{url}
\usepackage{hyperref}
\usepackage{todonotes}
\usepackage{enumitem}
\newtheorem{invar}{Invariant}
\usepackage{capt-of}
\usepackage{array, tabularx, makecell, colortbl}
\usepackage{newfloat}
\DeclareFloatingEnvironment[fileext=frm,placement={!ht},name=Algorithm]{myfloat}
\usepackage{multicol}
\usepackage{mdframed}
\usepackage{todonotes}
\usepackage{enumitem}
\usepackage{cite}
\def\checkmark{\tikz\fill[scale=0.4](0,.35) -- (.25,0) -- (1,.7) -- (.25,.15) -- cycle;}
\usepackage{pifont}
\newcommand{\xmark}{{\color{red}}\ding{55}}%
\usepackage{listings}
\usepackage{caption}
\usepackage{subfig}
\input{macros}

\author{
  Gaurav Bhardwaj \\
  Indian Institute of Technology \\
  Hyderabad.\\
  \texttt{cs19resch11003@iith.ac.in} \\
  \And
  Bapi Chatterjee \\
  Indraprastha Institute of Information Technology\\
  Delhi.\\
  \texttt{bapi@iiitd.ac.in} \\
  \And
  Abhinav Sharma \\
  Indraprastha Institute of Information Technology\\
  Delhi.\\
  \texttt{abhinav19006@iiitd.ac.in} \\
  \And
  Sathya Peri \\
  Indian Institute of Technology\\
  Hyderabad.\\
  \texttt{sathya\_p@cse.iith.ac.in}
  \And
   Siddharth Nayak \\
   Indraprastha Institute of Information Technology\\
   Delhi.\\
   \texttt{siddharth22128@iiitd.ac.in} \\
} 

\begin{document}
\title{Learned Lock-free Search Data Structures} 
\maketitle
\begin{abstract}
Non-blocking search data structures offer scalability with a progress guarantee on high-performance multi-core architectures. In the recent past, "learned queries" have gained remarkable attention. It refers to predicting the rank of a key computed by machine learning models trained to infer the cumulative distribution function of an ordered dataset. A line of works exhibits the superiority of learned queries over classical query algorithms. Yet, to our knowledge, no existing non-blocking search data structure employs them. In this paper, we introduce \textbf{Kanva}, a framework for learned non-blocking search. Kanva has an intuitive yet non-trivial design: traverse down a shallow hierarchy of lightweight linear models to reach the "non-blocking bins," which are dynamic ordered search structures. The proposed approach significantly outperforms the current state-of-the-art -- non-blocking interpolation search trees and elimination (a,b) trees -- in many workload and data distributions. Kanva is provably linearizable.
\end{abstract}

\section{Introduction} \label{sec:intro}
\input{Sections/introduction}

\section{Preliminaries} \label{sec:prelim}
\input{Sections/priliminaries}

\section{Kanva Data Structure Design} \label{sec:desing}
\input{Sections/KanvaDesign}

\subsection{Range Search}
\label{sec:rangesearch}
\input{Sections/rangsearch}

\section{Lock-free Asynchronous Model Fitting} \label{sec:lfmodel}
\input{Sections/lfmodel}

\section{Correctness of Algorithm} \label{sec:correct}
\input{Sections/correct}

\section{Evaluation} \label{sec:eval}
\input{Sections/Evaluation}
\section{Related Concurrent and Learned Indexes} \label{sec:review}
\input{Sections/BackgroundandMotivation}

\section{Conclusion} \label{sec:conclude}
\input{Sections/conclude}

\bibliography{citations}

\end{document}

%% file: macros.tex
\newcommand{\punt}[1]{}
\newcommand{\cmnt}[1]{}
\newcommand{\ignore}[1]{}
\newcommand{\vval}{\texttt{vValue}\xspace}
\newcommand{\bin}{\texttt{Bin}\xspace}
\newcommand{\obin}{\texttt{OLB}\xspace}
\newcommand{\tbin}{\texttt{TLB}\xspace}

\newcommand{\cbin}{child\_bin\xspace}
\newcommand{\nd}{\texttt{Node}\xspace}
\newcommand{\mnd}{\texttt{MNode}\xspace}
\newcommand{\knd}{\texttt{KNode}\xspace}
\newcommand{\model}{\texttt{Model}\xspace}

\definecolor{darkblue}{rgb}{0.0, 0.0, 0.55}

\newcounter{history}

\newcommand{\myparagraph}[1]{\indent\textbf{\large#1}}

\newcommand{\cas} {\texttt{\textbf{CAS}}\xspace}
\newcommand{\faa}{\texttt{\textbf{FAA}}\xspace}

\newcommand{\CAS}{compare-and-swap\xspace}

\newcommand{\lble} {linearizable\xspace}
\newcommand{\lbty} {linearizability\xspace}

\newcommand{\eabt} {Elim. $(a,b)$-Tree\xspace}
\newcommand{\lfabt} {LF $(a,b)$-Tree\xspace}
\newcommand{\cist} {C-IS-Tree\xspace}
\newcommand{\fdex} {FINEdex\xspace}
\newcommand{\xdex} {XIndex\xspace}
\newcommand{\kan} {Kanva\xspace}
\newcommand{\mrkan} {Kanva (MR)\xspace}

\newcommand{\seek}{\textsc{Seek}\xspace}
\newcommand{\rns} {\textsc{RangeSearch}\xspace}
\newcommand{\insertADT}{\textsc{Insert}\xspace}
\newcommand{\remove}{\textsc{Delete}\xspace}
\newcommand{\search}{\textsc{Search}\xspace}
\newcommand{\range}{\textsc{Range}\xspace}

\newcommand{\nul}{$null$\xspace}
\newcommand{\tru}{$True$\xspace}
\newcommand{\fal}{$False$\xspace}
\newcommand{\found}{\texttt{FOUND}\xspace}
\newcommand{\nfound}{\texttt{NFOUND}\xspace}
\newcommand{\maybe}{\texttt{MAYBE}\xspace}


%% file: Sections/introduction.tex
Dynamic hierarchical search data structures are the primary methods for predecessor and range queries. Today's multicore processors provide a natural hardware platform for the scalability of such algorithms with adaptability to streaming settings. The shared locks are the first approach to correctly translate a sequential data structure to a concurrent shared-memory system. However, irrespective of granularity, locks are prone to pitfalls such as deadlock, conveying, etc. By contrast, the \textit{lock-free (non-blocking) progress} ensures that some non-faulty threads finitely complete their operations. While a progress guarantee is desirable, consistency of concurrent operations is a necessity. The most popular consistency framework is \textit{\lbty} \cite{herlihy1990linearizability}, i.e., every concurrent operation takes effect at an atomic step between its invocation and return. In this work, we focus on the lock-free \lble implementation of hierarchical search data structures, oft-represented by search trees, for predecessor queries.

Starting from Ellen et al. \cite{ellen2010non} in PODC 2010, several lock-free binary search trees (BSTs) were proposed over the last decade: Howley and Jones \cite{howley2012non}, Natarajan and Mittal \cite{natarajan2014fast}, Chatterjee et al. \cite{chatterjee2014efficient}, Brown et al. \cite{brown2014general}. Arbel-Raviv et al. \cite{arbel2018getting}, Natarajan et al. \cite{natarajan2020feast}, and Brown et al. \cite{brown2022pathcas} further improved the results. While BSTs remained popular, during the same period, other lock-free hierarchical search structures were also proposed: k-ary search tree \cite{brown2011non}, B$^+$ tree \cite{braginsky+:SPAA:2012} and variants such as BW-tree \cite{wang2018building}, and skip-lists with batch updates \cite{kobus2022jiffy}. Further, the lock-free interpolation search tree (C-IST) by Brown et al. \cite{brown2020non} presented impressive empirical performance along with good theoretical guarantees (see Section \ref{sec:review}). Later, elimination $(a,b)$-tree \cite{srivastava2022elimination}, which is a lock-based variant of classical B$^+$ tree with variable (between $a$ and $b$) number of keys in nodes, claimed outperforming the then existing concurrent search trees. These works generally consider membership queries with concurrent updates in the data structures. The range queries mostly received generic approaches in the literature for inclusion into concurrent linked lists and BSTs: \cite{chatterjee2017lock}, \cite{arbel2018harnessing}, \cite{nelson2021bundled}, \cite{wei2021constant}, \cite{sheffi2022eemarq}.

Incidentally, the last decade also witnessed an unprecedented all-pervasive proliferation of machine learning (ML) techniques, influencing query algorithms. Kraska et al. \cite{kraska2018case} proposed a new perspective on search queries: obtain the approximate rank of a key using ML models trained to predict the cumulative distribution function of data. The approximation error would work as a tunable hyperparameter to define bounded proximity for the key's location. Kraska et al. \cite{kraska2018case} named the algorithm \textit{recursive model index (RMI)}, and the general approach a \textit{learned index}. RMI did not allow dynamic updates. The experiments of \cite{kraska2018case} showed that RMI performed up to 3x better than B+ trees for range queries.

Subsequent works on learned indexes incorporated dynamic updates, for example, ALEX \cite{Alex} and PGM-index \cite{ferragina2020pgm}, that use a hierarchy of learned linear regression models to predict the approximate key-rank. ALEX keeps gaps in the sorted data array for dynamic updates and splits them when needed. On the other hand, PGM-index uses a strategy similar to log-structured merge-trees (LSM-tree) \cite{o1996log} wherein an insert causing overflow at one data structure level is pushed to another level with a bigger storage capacity. A similar approach was also adopted by Radix-spline \cite{kipf2020radixspline}. 

Dynamic updates in learned indexes naturally motivated exploring their concurrent design. As the first approach, \xdex \cite{tang2020xindex} used locks on RMI of \cite{kraska2018case} with augmented buffers for each key segment to ingest dynamic updates. With such a design, a query searches for a key in a segment and the associated buffer; while the former uses a learned model, the latter employs the traditional comparison-based search. When the buffer is full, it is merged with its segment under a barrier. Thereon, the RMI is retrained.

The model retraining under barrier-based synchronization of \cite{tang2020xindex} was improved upon by \fdex \cite{li2021finedex}. They proposed a heuristically built shallower learned index fitted with low-depth B-trees, called bins, for ingesting dynamic updates. On reaching a storage threshold, the bins are replaced by local ML models under fine-grained locks. This allows \textit{two different} segments to receive updates and model retraining concurrently (which they seem to have claimed as non-blocking). However, at the data structure level, the progress guarantee is still blocking. The concurrent operations in \fdex are not \lble. Empirically, \fdex outperforms \xdex across various workloads.

Both \xdex and \fdex support range queries concurrent with updates and membership queries. Though \xdex discusses limited \lbty of update and membership operations, it is not clear if these operations are \lble in \fdex. It is important to note here that range queries in both \xdex \cite{tang2020xindex} and \fdex \cite{li2021finedex} are \textit{not \lble}.

This paper introduces \textbf{\kan}, a lock-free learned index structure that supports concurrent \lble membership, range and update operations. More specifically, 

\begin{enumerate}[noitemsep, topsep=0.2pt, leftmargin=*, nosep, nolistsep]
	\item we present a formal notion of search data structures (Section \ref{sec:prelim}) to review the existing concurrent and learned indexes under a common framework. (Section \ref{sec:review}).
	\item we describe the design of \kan, which integrates learned queries and lock-free synchronization techniques to enable fast \lble reads, updates, range search, and model retraining without any buffer, lock, or barrier. The core of our \lble range search is a non-trivial versioning approach for model arrays. (Section \ref{sec:desing}).
	\item unlike \fdex, our local model training is hyperparameter free; this is enabled by a clever use of exponential search for the last mile exploration. (Section \ref{sec:desing}).
    \item We extensively evaluate the presented algorithm. \kan outperforms state-of-the-art lock-free C-IST \cite{brown2020non} by up to two orders of magnitudes across the workload. It also significantly outperforms the elimination $(a,b)$ tree. Concerning learned rage queries, \kan significantly outperforms \fdex while offering \lbty. The code is available at \url{https://anonymous.4open.science/r/Kanva-CD83} (Section \ref{sec:eval}). 
    \item We discuss why the future of lock-free search structures should be learned. (Section \ref{sec:conclude}).
\end{enumerate}

%% file: Sections/priliminaries.tex
We consider \textbf{a Machine Learning Model}, often called \textbf{a model} for brevity, representing the learning outcome of an ML algorithm on a dataset. Formally, a model is a set of parameters $x$, which is implemented as a vector in $d$-dimensional real space: $x\in\mathbb{R}^d$.

We consider \textbf{an Abstract Data Type (ADT)} $\mathcal{A}$ that is a set of operations $\{\insertADT(K,V)$, ~$\remove(K)$, ~$\search(K)$,  $\range(K,r)\}$ on a key-value store $\mathcal{S} = (\mathcal{K, V})$, where $\mathcal{K}$ is a partially ordered set, also called the \textit{dataset} and $\mathcal{V}$ is a set of objects such that there is a bijection between $\mathcal{K}$ and $\mathcal{V}$. More specifically,
\begin{inparaenum}[(i)]
	\item An $\insertADT(K, V)$ inserts the key $K$ to $\mathcal{K}$ and the associated value $V$ to $\mathcal{V}$ and returns true if $K\notin\mathcal{K}$ otherwise if $K\in\mathcal{K}$ and $V\notin\mathcal{V}$, it inserts $V$ to $\mathcal{V}$ and returns true, and if $K\in\mathcal{K}\wedge V\in\mathcal{V}$ it returns false without any updates in either $\mathcal{K}$ or $\mathcal{V}$,
	\item A $\remove(K)$ deletes the key $K$ from $\mathcal{K}$ and its corresponding value $V\in\mathcal{V}$ if $K\in\mathcal{K}$;  if $K\notin\mathcal{K}$, it returns false without any modifications to either $\mathcal{K}$ or $\mathcal{V}$,
	\item A $\search(K)$ searches the key $K$ in $\mathcal{K}$ and return the corresponding $V\in\mathcal{V}$ if $K\in\mathcal{K}$; otherwise, it returns $null$, and
	\item A $\range(K,r)$ returns the set $V_R=\{V_i\}\subset\mathcal{V}$ corresponding to $K_R=\{K_i\}\subset\mathcal{K}$ such that $K\le K_i\le K+r$; if no such $K_i$ exists, it returns $null$.
\end{inparaenum}

We employ \textbf{a Search Data Structure} $\mathcal{D}$ to implement the ADT $\mathcal{A}$. For efficient implementation of ADT operations over $\mathcal{S}$, $\mathcal{D}$ is equipped with an \textbf{index} $\mathcal{I}$. A specific data point $R\in\mathcal{K}$ is designated as the \textit{root} of $\mathcal{D}$. \textbf{A Hierarchical Index} is a hierarchy of arithmetic or logical expressions to compute the \textbf{query path} from the root $R\in\mathcal{K}$ to the potential location of a query key $K\in\mathcal{K}$. Formally, it can be expressed as a finite ordered set $\mathcal{E}=\{e_i\}_{i=1}^r$, where each $e_i$ is \textit{an arithmetic or logical \textbf{expression}}. Update operations -- $\insertADT(K, V)$ and $\remove(K)$ modify $\mathcal{E}$ in a dynamic setting. 
\input{Diagram/Design}

\textbf{A Comparison-based Index}, often referred to as \textbf{traditional} or \textit{classical index}, primarily uses logical operations $<,~\le,~=,~\ge,~>,$ to determine the query path. Each $e_i$ comprises one or more logical operations along with some data. For example, in a binary search tree, each internal node, as a representative $e_i$, will contain a key $x$ along with logical operations $<,~\ge$.  
\input{Diagram/Design1}

\textbf{A Learned Index} uses arithmetic expressions that are machine learning models to compute the query path. In essence, the models $\mathcal{E}=\{e_i\}_{i=1}^r$ together approximate the cumulative distribution function (CDF) $F$ of the set $\mathcal{K}$. If $\tilde{F}$ is an approximation to $F$, then the potential position of a key $K\in\mathcal{K}$ will belong to the interval
\[[\tilde{F}(K)\times |\mathcal{K}|-\epsilon,~ \tilde{F}(K)\times |\mathcal{K}| + \epsilon],\]

where $\epsilon$ is a tunable hyperparameter related to the approximation error. An important difference between a comparison-based index and a learned index, as shown in Figure \ref{fig:index}, is that a query path from root $R\in\mathcal{K}$ to a key $K\in\mathcal{K}$ is unique in the former, whereas in the latter, in some designs, for example in RMI \cite{kraska2018case}, there can be multiple query paths depending on the inference by different models in the hierarchy.

\input{table/Scalability}

\textbf{The Shared Memory System} considered in this work supports atomic \texttt{read}, \texttt{write}, and \texttt{\cas} (\CAS) primitives and comprises a finite set of threads $\{t_i\}_{i=1}^p$. 

\textbf{A \lble Lock-free Data Structure} implements the ADT operations in a shared memory system such that in every concurrent execution of operations, the effect of each of them will be visible to a user between their invocation and response to ensure \lbty. Furthermore, in such executions, at least one non-faulty operation will finish in a finite number of steps, irrespective of the behavior of other concurrent operations, to ensure \emph{lock-freedom}.

\textbf{A \lble Learned Lock-free Data Structure} is an {updatable} ordered set of expressions $\mathcal{E}=\{e_i\}_{i=1}^r$, where some or all of $e_i$s are machine learning models. As implemented on a shared memory system, the ADT operations in such a data structure are \lble and lock-free.

Table \ref{Table:scalability} summarizes relevant concurrent, learned, and concurrent \& learned indexes.

%% file: Diagram/Design.tex
\begin{figure}[h!]
\captionsetup[subfigure]{justification=centering}
\centering
  \subfloat[]{  \includegraphics[width=0.48\linewidth]{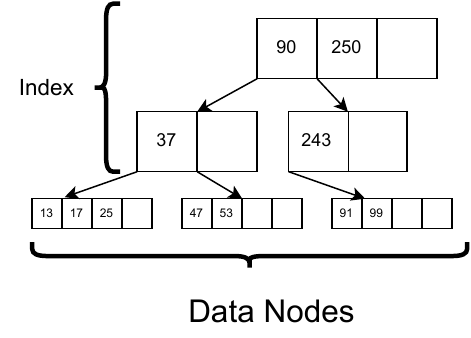}}
   \subfloat[]{ 
  \includegraphics[width=0.48\linewidth]{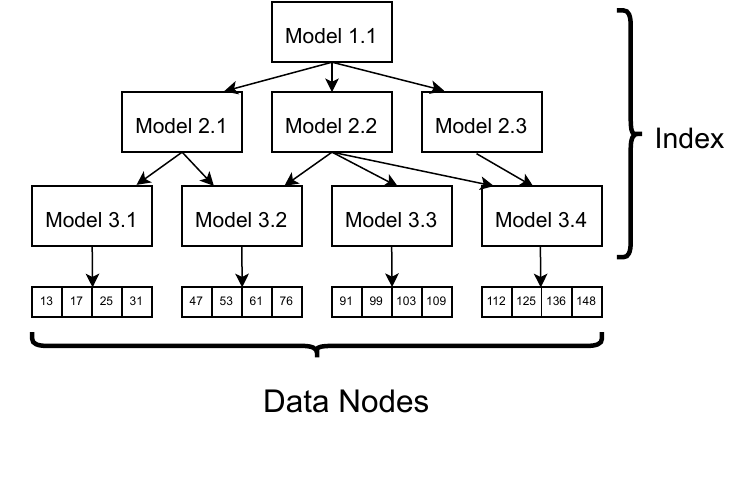}}
\vspace{-5mm}
\caption{(a) Traditional Index (b) Learned Index}
\label{fig:index}
\end{figure}

%% file: Diagram/Design1.tex
\begin{figure*}[ht!]
\centering
\begin{tabular}{c c c c}
\subfloat[] { \includegraphics[width = 0.23 \linewidth]{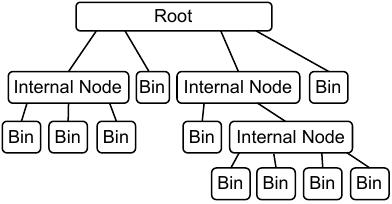} }
&
\subfloat[] { \includegraphics[width = 0.23 \linewidth]{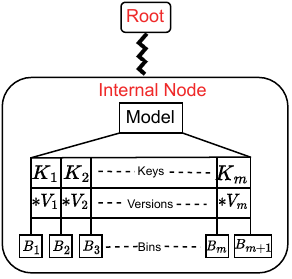} }
&
\subfloat[] { \includegraphics[width = 0.23 \linewidth]{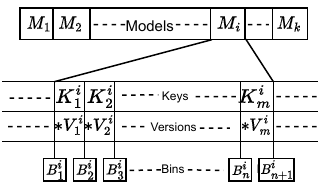} }
&
\subfloat[] { \includegraphics[width = 0.18\linewidth]{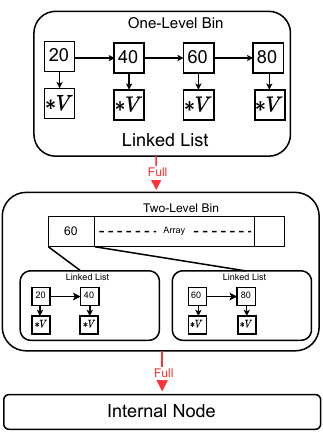}
\label{fig:Lifecycle}
}
\end{tabular}
\caption{Kanva Design structure: (a) Shallow Hierarchy (b) An internal node (c) The root node (d) The lifecycle of a Bin}
\label{fig:kanva-structure}
\end{figure*}

%% file: table/Scalability.tex
\begin{table}[htbp]
	\tiny
\centering
\begin{tabular}{|p{0.125\columnwidth}|p{0.06\columnwidth}|p{0.065\columnwidth}|p{0.04\columnwidth}|p{0.025\columnwidth}p{0.025\columnwidth}p{0.025\columnwidth}p{0.03\columnwidth}|p{0.1\columnwidth}|}
\hline
\multirow{2}{*}{\textbf{Algorithms}} &
  \multirow{2}{=}{Learned} &
  \multirow{2}{=}{Updates} &
  \multirow{2}{=}{Data Sorted} &
  \multicolumn{4}{|c|}{Concurrent Insert/Remove/Search} &
 \multirow{2}{=}{Range Queries} \\ \cline{5-8}
 &
   &
   &
   &
  \multicolumn{1}{l|}{Write} &
  \multicolumn{1}{l|}{Retrain} &
  \multicolumn{1}{l|}{Lock-free} &
  \multicolumn{1}{l|}{Linearizable}    &  \\ \hline
\cist ~\cite{brown2020non}       & \multicolumn{1}{c|}\xmark  & \multicolumn{1}{c|}\checkmark                 & \multicolumn{1}{c|}\checkmark  & \multicolumn{1}{c|}{\checkmark} & \multicolumn{1}{c|}{\xmark}   & \multicolumn{1}{c|}{\checkmark}    & \checkmark & Not present \\ \hline
\lfabt ~\cite{TrevorBrown:PhDThesis}       & \multicolumn{1}{c|}\xmark  & \multicolumn{1}{c|}\checkmark                 & \multicolumn{1}{c|}\checkmark  & \multicolumn{1}{c|}{\checkmark} & \multicolumn{1}{c|}{\xmark}   & \multicolumn{1}{c|}{\checkmark}    & \checkmark & Not present\\ \hline
\eabt ~\cite{srivastava2022elimination}       & \multicolumn{1}{c|}\xmark  & \multicolumn{1}{c|}\checkmark                 & \multicolumn{1}{c|}\checkmark  & \multicolumn{1}{c|}{\checkmark} & \multicolumn{1}{c|}{\xmark}   & \multicolumn{1}{c|}{\xmark}    & \checkmark & Not present \\ \hline
RMI ~\cite{kraska2018case}       & \multicolumn{1}{c|}\checkmark  & \multicolumn{1}{c|}\xmark                 & \multicolumn{1}{c|}\checkmark  & \multicolumn{4}{c|}{\xmark} & Sequential\\ \hline
FITing Tree ~\cite{galakatos2019fiting}       & \multicolumn{1}{c|}\checkmark   & \multicolumn{1}{c|}\checkmark                  & \multicolumn{1}{c|}\xmark & \multicolumn{4}{c|}{\xmark}   & Sequential\\ \hline
ALEX ~\cite{Alex}         & \multicolumn{1}{c|}\checkmark         & \multicolumn{1}{c|}\checkmark                  & \multicolumn{1}{c|}\xmark & \multicolumn{4}{c|}{\xmark}  & Sequential\\ \hline
PGM-Index ~\cite{ferragina2020pgm}          & \multicolumn{1}{c|}\checkmark   & \multicolumn{1}{c|}\checkmark                  & \multicolumn{1}{c|}\xmark  & \multicolumn{4}{c|}{\xmark}   & Sequential \\ \hline
XIndex ~\cite{tang2020xindex}        & \multicolumn{1}{c|}\checkmark        & \multicolumn{1}{c|}\checkmark                  & \multicolumn{1}{c|}\xmark & \multicolumn{1}{c|}{\checkmark}  & \multicolumn{1}{c|}{\checkmark}    & \multicolumn{1}{c|}{\xmark} & \checkmark    & Blocking Not Linearizable\\ \hline
\fdex ~\cite{li2021finedex}        & \multicolumn{1}{c|}\checkmark       & \multicolumn{1}{c|}\checkmark                  & \multicolumn{1}{c|}\checkmark  & \multicolumn{1}{c|}{\checkmark}  & \multicolumn{1}{c|}{\checkmark}    & \multicolumn{1}{c|}{\xmark}   & \xmark & Blocking Not Linearizable \\ \hline
\textbf{\normalsize Kanva (ours)}       & \multicolumn{1}{c|}\checkmark         & \multicolumn{1}{c|}\checkmark                  & \multicolumn{1}{c|}\checkmark  & \multicolumn{1}{c|}{\checkmark}  & \multicolumn{1}{c|}{\checkmark}    & \multicolumn{1}{c|}{\checkmark}   & \checkmark   & \textbf{Lock-free Linearizable}\\ \hline
\end{tabular}%
\caption{Concurrent and Learned Indexes}
\label{Table:scalability}

\end{table}

%% file: Sections/KanvaDesign.tex
\subsection{Layout and Operations basics}

Structurally, \kan capitalizes on \fdex \cite{li2021finedex} to implement a \lble lock-free dynamic key-value store. \kan implements an ADT that we presented in the last section. Our approach for \lble range search requires pairing the values with their update timestamps drawing from \cite{wei2021constant}. For this purpose, we associate a list of objects packing a value and its corresponding update timestamp to a key. We refer to this object as \textit{\vval} or versioned-value. The design and evolution of \kan is shown in Figure \ref{fig:kanva-structure}.

\textbf{Models:} We use \textbf{linear regression models} in \kan nodes to approximate the ranks of the keys as \[rank=a\times key + b + \epsilon.\] We apply a simple lock-free linear model fitting approach as described in Section \ref{sec:lfmodel}. Given a dataset as a sorted array of keys, models essentially approximate the CDF of key segments of the dataset. Thus, a set of models approximate the CDF of the dataset in a \textit{piecewise linear} fashion. A model is represented as a packet of parameters $\{a,b,\epsilon\}$.

\textbf{Hierarchical structure} of \kan as shown in Figure \ref{fig:kanva-structure}(a) is shallow and unbalanced. It comprises \textit{{internal nodes}} and \textit{{Bins}}. Both internal nodes and bins contain multiple key-value pairs and are updatable. 

\textbf{Internal nodes} consist of a sorted array of keys and an array of \vval pointers, see Figure \ref{fig:kanva-structure}(b). An internal node \textit{other than the root} is equipped with a single model. Internal nodes are also updatable: inserting a key-value pair with a new value updates the latest \vval object associated with the key, and a deleted key can be re-inserted with a new associated value. Each internal node other than the root is roughly equal in size, which is determined by the size threshold of a bin, which we specify below. An internal node containing $m$ keys has $m+1$ pointers to bins associated with it following the order of the keys. The bin pointers are $null$ until a key-value pair is stored.

\ignore{wiFor the  Unlike \fdex \cite{li2021finedex}, we apply a single pass over data to build the models and compute the error hyperparameter $\epsilon$. 
 This approach is beneficial in a lock-free concurrent setting with dynamic updates also in the internal nodes.  
}
\textbf{Bins} of \kan ingest the insertion and deletion of key-value pairs. A bin is structurally different from an external node of a classic B$+$ tree. Therefore, we named them so. We initialize a bin as a sorted linked list called a \textbf{one-level bin}. As the number of key-value pairs in a one-level bin reaches its size threshold, it is replaced with a \textbf{two-level bin}. A two-level bin has an array for indexing and multiple one-level bin pointers, which can further store key-value pairs. As the number of keys in a two-level bin reaches its size threshold, it is replaced with an internal node containing its key-value pairs. The lifecycle of a bin is shown in Figure \ref{fig:kanva-structure}(d). 

\textbf{The root} of \kan is a unique internal node that contains the entire dataset at initialization; see Figure \ref{fig:kanva-structure}(c). It includes an array of models for a piecewise linear fitting to the entirety of the initially given dataset. Using a model fitting scheme described in Section \ref{sec:lfmodel}, we obtain key segments with an identical error-bound $\epsilon$ for linear regression models as mentioned above. 
Initially, $n+1$ bin pointers are available in the root to ingest the updates, where $n$ is the size of the initially given dataset. 

\myparagraph{Traversals} in \kan are powered by the ML models. With multiple key segments in the root, we start by searching a query key over the set of endpoints, often applying a binary search. Having searched the appropriate segment, we search the key in the segment enabled by its associated model. Thus, the approximate location of the key is fixed. After that, a binary search in a bounded vicinity, determined by the regression error $\epsilon$, is performed to find the key's exact (possible) location. We perform an exponential search in an internal node other than the root to find the precise location. Exponential search does not require computing $\epsilon$, which is beneficial in a lock-free setting, as we describe in Section \ref{sec:lfmodel}. 
An operation only travels to the next level if the query key is not found at an internal node. Thus, a shallow hierarchy substantially reduces the lookup cost in \kan. A linear search is performed in a one-level bin, whereas, in a two-level bin, a linear search follows a binary search to determine the appropriate one-level bin. 

\subsection{Data types}
\input{Sections/app}

The objects to implement \kan are described in Algorithm \ref{fig::Model}.  
Both internal nodes and bins in \kan are instances of class \nd. An internal node instantiates the class \mnd (Modelled Node). A model is represented by a set of parameters as implemented by the class \model. An \mnd object encapsulates a \model array \texttt{models[]} fitted to the keys that it contains. The associated list of \vval objects is linked by their head pointers stored in the array \texttt{versions}. 

The class \bin maintains its size threshold and a boolean to indicate if it is a two-level bin. A \bin class is further inherited by the classes \tbin and \obin of which instances implement two-level bins and one-level bins, respectively. The keys and associated links to \vval objects are encapsulated by class \knd. The \vval objects instantiate the class \vval. 

An instance of \kan contains an atomic integer \texttt{Timestamp} that helps in versioning the updates to enable \lble range search as we discuss in Section \ref{sec:rangesearch}.

\subsection{Lock-free \insertADT, \remove, and \search}
\textbf{\seek} method implements traversals in \kan as described above, which every operation calls at the invocation; see Algorithm \ref{alg:code-traverse}. \seek calls the method $searchInMNode$ to determine the index of the key array where the key could be potentially located. If the key is not located in the \mnd, $searchInMNode$ returns the index of the \texttt{children} array, where seek should be directed to following the sorted order. As a requirement, \seek returns a packet of \mnd, the index of a child pointer or the index of the key if located, and an enumeration -- $\{\found, \nfound, \maybe\}$ -- of the status of locating the key. 


\input{algorithm/code-traverse}


%

\input{algorithm/read1}

\myparagraph{\insertADT} is given in lines \ref{insert} to \ref{insert_end} in Algorithm \ref{fig:code}. 
It performs traversal at line \ref{ins:find_idx} to find the node where the insert can happen. If the key is found in an \mnd, in that case, it attempts to update the corresponding value using an atomic compare-and-swap (\cas) in the method \textsc{writeValue}, which repeatedly attempts to update the value until a \cas succeeds. 


If the key doesn't exist in an \mnd, the operation will try inserting it in an appropriate $Bin$. If the $Bin$ doesn't exist, a new $Bin$ is created at line \ref{newlevelbinstart} and atomically inserted using a \cas at line \ref{levelcas}. If the $CAS$ operation fails at line \ref{levelcas}, then some other thread must have added the $Bin$, and therefore the operation is retried. If the $Bin$ exists and has reached the threshold or is already undergoing transformation to a \mnd, it will engage in helping via method $\textsc{helpMakeModel}$ at lines \ref{ins:makemodel} or \ref{ins:makemodel1}. After any such possible help, insertion is attempted in a bin using method $insertBin$, whose result is returned finally.

\myparagraph{\remove} is given in lines \ref{del:start} to \ref{del:end}. \kan doesn't perform the physical deletion of a key. Instead, it marks the key as deleted by updating its value to \nul. In that way, a \remove operations works exactly like an \insertADT operation to update the corresponding value of a present key. If the key is not found it returns \fal. 




\myparagraph{\search} operations, given in lines \ref{srch:start} to \ref{search:srch_bin} in Algorithm \ref{fig:code}, similarly starts with calling \texttt{Seek} in line \ref{srch:seek}. In case \seek returns \found or \maybe, and the fact that for the deleted keys, the corresponding values are \nul, it is straightforward for a \search operation to implement its ADT definition. If \seek returns \nfound, \search returns \nul.

\myparagraph{Helping} is performed in the method \textsc{helpMakeModel} as given in Algorithm \ref{alg:help}. Essentially, helping is required when a bin reaches its threshold. For both \obin and \tbin, we start with freezing its \knd{s}. For this purpose, we adopt the popular bit-stealing technique \cite{Harris:NBList:DISC:2001} to use one of the unused bits of their next pointers. On observing a frozen bin, $insertBin$ and $deleteBin$ operations result into indicating that a \insertADT and \remove operation need to engage in helping. Once an \obin is frozen, a \tbin replaces it appropriately. For a frozen \tbin, a key array is created of the keys collected from all its \obin{s}. After that, we perform a lock-free model training on this key array as described in Section \ref{sec:lfmodel} using the method \textsc{makeModel}. The created model and key array make a new \mnd; see line \ref{help:newmnode}.

\input{algorithm/help_make_model.tex}

%% file: Sections/app.tex
\begin{figure}[h!]
\begin{minipage}[t]{.49\columnwidth}
\begin{lstlisting}[basicstyle=\scriptsize]
class Node{ }
class Model{
    float parameters[];
}
class MNode inherits Node{
    Model models[];
    k_type keys[];
    vValue* versions[];
    Node* children[];
}
class Bin inherits Node{
    int size;
    bool isOneLevel;
}
class OLB inherits Bin{
	KNode* head;
}
\end{lstlisting}
\end{minipage}
\begin{minipage}[t]{.49\columnwidth}
   \begin{lstlisting}[basicstyle=\scriptsize] 
class TLB inherits Bin{	
    k_type keys[];
    KNode* children[];
}
class KNode{
    k_type item;
    vValue* version;
    KNode* next;
}
class vValue{
    v_type val;
    int ts;
    vValue* vnext;
}
MNode root = BuildKanva(Dataset);
AtomicInt Timestamp = 0;
\end{lstlisting}
\end{minipage}
\captionof{myfloat}{Kanva: Data Structure Objects}
\label{fig::Model}
\end{figure}

%% file: algorithm/code-traverse.tex
\begin{figure}[h]
		\begin{algorithmic}[1]
			\renewcommand{\algorithmicprocedure}{}
			\label{Read}
			\scriptsize
			\Procedure{\textbf{Seek}}{key, node}
			\State{$ix \gets searchInMNode(node, key);$}
			\label{find_contains}
			\If{($node.keys[ix] = key$)}
			\State{\texttt{\textbf{return (node, ix, \found)}}}
			\label{find_return}  
		\Else
		\State{$childNode \gets node.children[ix+1]$}
		\If{($childNode = null$)}
			\State{\texttt{\textbf{return (node, ix, \nfound)}}}         \label{find_return_false}  
		\ElsIf{$(type(childNode)=\bin)$}
			\State{\texttt{\textbf{return (node, ix, \maybe)}}}  
			\label{search:return_-1}
		\Else
		\State{\texttt{\textbf{return}} $\textsc{Seek}(key, childNode)$}
		\label{search:srch_node}
		
		\EndIf
		\EndIf
		\EndProcedure
		\algstore{seek}
	\end{algorithmic}
\captionof{myfloat}{Traversals in \kan.}
\label{alg:code-traverse}
\end{figure}

%% file: algorithm/read1.tex
\begin{figure}[h]
\begin{multicols}{2}
\begin{algorithmic}[1]
\renewcommand{\algorithmicprocedure}{}
\scriptsize
\algrestore{seek}
    \Procedure{\textbf{\insertADT}(key, value)}{}
    \label{insert}
     \State{\texttt{retry:}}
   	\State{($node, ix, status$) $\gets$ \textsc{Seek}($key, Root$)}
   	\label{ins:find_idx}
   	\If{$(status=\found)$}
   		\State{\texttt{\textbf{return}}  \textsc{writeValue}(node, ix, \textsc{value})}
   		\label{insert:value}
\ElsIf{$(status=\nfound)$}   
       \State{$newBin \gets$ new Bin(\textsc{key, value})}
       \label{newlevelbinstart}
       \label{newlvwlbinend}
       \If{$node.children[ix].\texttt{\textbf{CAS}}(\nul,$ $newBin)$}
       \State{\texttt{\textbf{return}} \tru}  
       \label{levelcas}
       \Else \texttt{ \textbf{goto}} retry
       \EndIf
       \label{insertretry1}
    \Else
        \State{$bin \gets node.children[ix]$}
        \If{($bin.size \geq threshold$)}
       \State{$\textsc{helpMakeModel}(node, ix, bin)$}
       \label{ins:makemodel}
       \State{\texttt{\textbf{goto}} retry}
        \Else
        \State{res $\gets$ $bin.insertBin(\textsc{key,value})$}
            \If{($res = \texttt{underMakeModel}$)}
             \State{$\textsc{helpMakeModel}(node, ix, bin)$}
             \label{ins:makemodel1}
            \State{\texttt{\textbf{goto}} retry}
            \Else {\texttt{\textbf{ return} res}}\label{insert_end}
            \EndIf
        \EndIf
    \EndIf

\EndProcedure
\algstore{ins}
\end{algorithmic}
\vspace{-12pt}
\rule{0.49\textwidth}{0.1pt}
\begin{algorithmic}
    \renewcommand{\algorithmicprocedure}{}
     \label{Alg::delete}
\scriptsize
    \algrestore{ins}
    \Procedure{\textbf{\remove}(key)}{}
    \label{del:start}
	\State{retry:}
	\State{$(node, ix, status) \gets \textsc{Seek}(key,Root)$}
	\If{($status = \found)$}
    \If{(\textsc{readValue}(node.versions[ix] $\neq$ \nul))}
     \State{\texttt{\textbf{return}} \textsc{\textsc{writeValue}}(node, ix, \nul)}
    \label{delete:value}
    \Else{ \texttt{\textbf{return}} false}
    \EndIf
\ElsIf{($status = \nfound$)}
   	\State{\textbf{return} false}
\Else
    \State{$bin \gets node.children[ix]$}
        \State{$res \gets bin.deleteBin(key)$}
         \label{del:delbin}
        \If{($res = \texttt{underMakeModel}$)}
       \State{$\textsc{helpMakeModel}(node, ix, bin)$ \State{\texttt{\textbf{goto}} retry}}
        \Else{\texttt{\textbf{ return}} res}
        \EndIf
        \EndIf
            \label{del:end}
\EndProcedure 
\algstore{del}
\end{algorithmic}
\vspace{-12pt}
\rule{0.49\textwidth}{0.1pt}
\begin{algorithmic}
\renewcommand{\algorithmicprocedure}{}
\scriptsize
\algrestore{del}
\Procedure{\textbf{\search}}{key}
\label{srch:start}
\State{($node, ix, status$) $\gets$ \textsc{Seek}($key, Root$)}
\label{srch:seek}
\If{$(status = \found)$}
\State{\textbf{return} \textsc{readValue} ($node.versions[ix]$)}
\ElsIf{($status = \nfound$)}
\label{srch:levelnode}
\State{\textbf{return} $null;$}
\Else
\State{$bin \gets node.children[ix]$}
\State {$binNode\gets searchBin(bin,key)$}
\label{search:srch_bin}
\If{($binNode.key = key$)} 
\State{\textbf{return} \textsc{readValue}($binNode.version$)}
\Else
\State{\textbf{return} $null;$}
\EndIf
\label{seek:srch_bin}
\EndIf
\label{search-end}
\EndProcedure
\algstore{srch}
\end{algorithmic}
\vspace{-10pt}
\rule{0.49\textwidth}{0.1pt}
\begin{algorithmic}
\scriptsize
 \algrestore{srch}
 \renewcommand{\algorithmicprocedure}{}
 \Procedure{\textbf{initTS}}{vValue$\ast$ ver}
 \If{$ver.ts = -1$}
 \State{$\textbf{CAS}(ver.ts, -1, timestamp)$}
 \EndIf
 \EndProcedure
 \algstore{inits-1}
\end{algorithmic}
\vspace{-10pt}
\rule{0.49\textwidth}{0.1pt}
\begin{algorithmic}
\scriptsize
 \algrestore{inits-1}
 \renewcommand{\algorithmicprocedure}{}
 \Procedure{\textbf{readValue}}{vValue$\ast$ ver}
 \State{$\textsc{initTS}(ver)$}
 \State{\texttt{\textbf{ return}} ver.val}
 
 \EndProcedure
 \algstore{readvalue}
 \end{algorithmic}
\vspace{-10pt}
\rule{0.49\textwidth}{0.1pt}
\begin{algorithmic}
\scriptsize
 \algrestore{readvalue}
 \renewcommand{\algorithmicprocedure}{}
 \Procedure{\textbf{writeValue}}{node, ix, val}
 \While{(true)}
 \State{$currVhead \gets node.versions[ix]$}
 \State{$\textsc{initTS}(currVhead)$}
 \If{($currVhead.val = val$)} \texttt{\textbf{ return}} false
 \EndIf
 \State{$newVer \gets new vValue(val)$}
 \State{$newver.vnext = currVhead$}
 \If{(\textbf{CAS}($node.versions[ix], currVhead, newver$))}
 \State{\textsc{initTS}(newver)}
 \State{\texttt{\textbf{ return}} true}
 \EndIf 
 \EndWhile
 \EndProcedure
 \algstore{writevalue}
 \end{algorithmic}
\end{multicols}
\captionof{myfloat}{\insertADT, \remove, and \search operations.}
\label{fig:code}
\end{figure}

%% file: algorithm/help_make_model.tex
\begin{figure}
 \begin{algorithmic}[1]  
 \scriptsize
 \algrestore{writevalue}
 \renewcommand{\algorithmicprocedure}{\textbf{Method}}
\Procedure{helpMakeModel}{node, ix, bin}
\label{help_model}
        \State{$binKeys, binVersions, model \gets$  $bin.\textsc{make}\textsc{Model}()$}
       \State{$newNode \gets$ new MNode($binKeys,$ $binVersions, model$)}
       \label{help:newmnode}
       \State{$node.children[ix].\texttt{\textbf{CAS}}(bin,$ $newNode)$}
       \label{help:bin_cas}
       \State{\texttt{\textbf{ return}}}
\EndProcedure
\algstore{help}
\end{algorithmic}
\vspace{-5pt}
\rule{0.49\textwidth}{0.1pt}
\begin{algorithmic}
    \scriptsize
 \algrestore{help}
 \renewcommand{\algorithmicprocedure}{\textbf{Method}}

\Procedure{makeModel}{}
\State{$Keys$, $Versions$:= Collect all the keys and versions after freezing each node one by one.}

\State{$Model$:= $makeModel(Keys, Versions)$
\State{\texttt{\textbf{ return}} $Keys, Versions, Model$}}
\label{helpmodel:end}
\EndProcedure
\algstore{model}
\end{algorithmic}
\captionof{myfloat}{\textsc{helpMakeModel} and  \textsc{makeModel}}
\label{alg:help}
\end{figure}

%% file: Sections/rangsearch.tex
 \input{algorithm/rangesearch}
The \lble range search in \kan, as given in Algorithm \ref{alg:rangesrch},  uses the versioning approach of Wei et al.\cite{wei2021constant}. The \insertADT and \remove methods discussed above developed the ground for \range. As a key-value pair is inserted, deleted, or updated, the value gets its corresponding timestamp. With every key, the head of the associated \vval pointer points to the \vval object with the highest timestamp; see method \textsc{writeValue} in Algorithm \ref{fig:code}. 

See line \ref{line:ts}: after reading the atomic integer \texttt{Timestamp}, the method \textsc{readTimestamp} atomically increments it by 1. A range search will use this timestamp to collect the key-value pairs that have timestamps at most the one that it started with. Accordingly, its linearization point is decided as we discuss in Section \ref{sec:correct}. 
\rns calls the method \textsc{Scan} to recursively collect the target key-value pairs based on the timestamp. It is pertinent to underscore here the approach of fine-grained versioning of key-value pairs, even in the `fat' internal nodes in \kan is beneficial to the implemented \lble range search, which is not possible in lock-free ISTree \cite{brown2020non}, wherein they replace the internal nodes using double-compare-single-swap (DCSS).

%% file: algorithm/rangesearch.tex
\begin{figure}[h!]
\begin{multicols}{2}
 \begin{algorithmic}[1]  
 \scriptsize
 \algrestore{model}
 \renewcommand{\algorithmicprocedure}{}
\Procedure{\textbf{RangeSearch}}{key, range, result}
\label{alg:range_search}
\State{$ts \gets \textsc{readTimestamp()}$}
\label{line:ts}
\State{result = \nul}
\State{$\textsc{Scan}(root, key, range, result, ts)$}
\State{\texttt{\textbf{ return}} result}
 \EndProcedure
 \algstore{rs}
 \end{algorithmic}
 \vspace{-10pt}
\rule{0.49\textwidth}{0.1pt}
 \begin{algorithmic}
 \scriptsize
 \algrestore{rs}
 \renewcommand{\algorithmicprocedure}{}
\Procedure{\textbf{Scan}}{node, key, range, result, ts}
\label{Scan}
        \State{$ix \gets node.find\_idx(key)$}
        \While{$((ix \leq node.keys.size) \bigwedge (range > 0))$}
        \If{($node.keys[ix] \geq key$)}
        \State{$val = read(node.versions[ix], ts)$}
        \If{($val \neq \nul$)}
        \State{result.push\_back($node.keys[ix], val$)}
        \State{$range \gets range - 1$}
        \EndIf
        \EndIf
        \State{$node \gets node.children[ix]$}
        \If{$(type(node) = MNode)$}
        \State{$\textsc{Scan}(node, key, range, $ $result, ts)$}
        \ElsIf{$(type(node) = Bin)$}
        \State{$node.scanBin(key, range,result,$ $ts)$}
        \EndIf
        \State{$ix++$}
        \EndWhile
        \State{\textbf{return}}
\EndProcedure

\algstore{scan}
\end{algorithmic}
 \vspace{-10pt}
\rule{0.49\textwidth}{0.1pt}
\begin{algorithmic}
     \scriptsize
 \algrestore{scan}
 \renewcommand{\algorithmicprocedure}{} 
\Procedure{\textbf{readTimestamp}()}{}
\State{$ts \gets \texttt{Timestamp}$}
\State{$\textbf{CAS}(\texttt{Timestamp}, ts, ts+1)$}
\State{\texttt{\textbf{ return}} $ts$}
\EndProcedure
\algstore{takets}
\end{algorithmic}
 \vspace{-10pt}
\rule{0.49\textwidth}{0.1pt}
 
  
\begin{algorithmic}
\scriptsize
 \algrestore{takets}
 \renewcommand{\algorithmicprocedure}{}
 \Procedure{\textbf{readValue}}{vValue$\ast$ ver, ts}
 \State{$\textsc{initTS}(node)$}
 \While{($ver \bigwedge ver.ts > ts$)}
   \State{$ver \gets ver.vnext$}
  \EndWhile
  \If{($ver$)} (\texttt{\textbf{ return}} $ver.val$)
\EndIf
  \State{\texttt{\textbf{ return}} $tombstone$}
 \EndProcedure
 \end{algorithmic}
\end{multicols}
\captionof{myfloat}{Pseudocode of \textsc{RangeSearch}}
\label{alg:rangesrch}
\end{figure}

%% file: Sections/lfmodel.tex
In this section, we describe the choice of a model and its training algorithm. 
The experimental setup and other platform configurations are described in the next section, which are used for the experimental analysis that we present here.

\myparagraph{Lock-free regression:} We explored several options for computing a regression model. Firstly, as different threads collect the keys, they synchronize at the same array of entities ($\sum x$, $\sum y$, $\sum x^2$, $\sum y^2$, $\sum xy$, where $x$ represents a key and $y$ represents its rank in the array \texttt{keys[]}) for computing the regression co-efficient in a lock-free manner. This can be done via \cas as well as via fetch-and-add (\faa) primitives. We also explored the celebrated approach of Hogwild! \cite{recht2011hogwild} for training a regression model via gradient updates; however, it is essentially a multi-pass scheme and performs poorly if only one pass over the dataset is applied. Finally, we allocated different memory for selected keys and regression coefficient by different threads, and we selected one of the resultant ones after its completion in a single-step lock-free synchronization. The regression error and corresponding latency results are given in Table \ref{table:collect}. 
\input{table/Modelcount}

Clearly, the last one works best to select our threshold, which we experimentally traded off against the overall performance. Note that our linear model fitting approach is one pass, resulting in non-identical $\epsilon$ across the nodes.

\myparagraph{Bin Structure Selection:} For overall performance of the data structure, we also explored the choices of bins with different index choices; we considered (a) non-blocking skip-lists \cite{herlihy2006provably} (b) non-blocking two-level bins as described in Section \ref{sec:desing} (c) a variant of our introduced two-level bin, where we have linked lists replaced with arrays in the hope of getting advantage of better cache behaviour.  For (c), we needed to reallocate memory for an array of size 32 for every update. 

\input{Diagram/PGM}
\myparagraph{Index Selection:} Figure \ref{fig:PGM} shows the comparative throughput for different combinations of model fitting algorithms and non-blocking bin structures. We considered two different model fitting algorithms PGM \cite{ferragina2020pgm} and PLA of \cite{xie2014maximum}, for our lock-free approach to build on. PLA of \cite{xie2014maximum} is a heuristic approach wherein a regression model is created using a number of keys. Suppose the residual error is found to be more than that of the predetermined bound of the approximation error. In that case, the regression model computation is reattempted over a smaller set of keys. While PGM is a single-pass scheme, the latter is a multi-pass approach. Still, we found that in a concurrent setting, a directly computed regression model works better than a geometric linear model fitting approach of \cite{ferragina2020pgm}.

Our proposed two-level non-blocking bins with a lock-free model fitting algorithm offer the maximum throughput as shown in Figure \ref{fig:PGM}. This can be understood as memory reallocation cost overshadowing marginal cache locality optimization. Furthermore, it also establishes the superiority of the chosen model fitting algorithm.  We recorded the cardinality of the hierarchy, i.e. the total number of comparison-based or model-based nodes created in each method. The numbers are presented in Table \ref{table:nodes}. It can be seen that the concurrent learned indexes have a much smaller hierarchy than concurrent classical ones.

\ignore{\myparagraph{Building Latency:} The time taken to build the structures based on different indexes are presented in the Appendix in Table \ref{time}. We can see that the single pass model based learned indexes take significantly lesser time to build the search structure. Conventional tree structures like \lfabt and \cist create a large number of nodes, and the indexing time is also very high. \cist needs to calculate the parameters for interpolation search every time it rebalances, which makes it costlier in terms of time as the number of data items increases. This explains the better performance of learned index schemes regarding memory and time. Moreover, single-pass techniques give the optimal number of models and take less time to prepare the model.

}


\ignore{In our experiments, we have used LPA for models. In the case of bins, we have used a non-blocking linearizable two level bin. We have compared Kanva with a different state-of-the-art learned index as well as conventional schemes in all different types of distribution and real-time data such as amazon, Facebook, OSM and Wikipedia which represents two-dimensional data into one-dimensional space. We have also compared some data distributions such as Uniform and Log-Normal. 
}


%% file: table/Modelcount.tex
\begin{table}[h]
	\resizebox{\columnwidth}{!}{%
		\begin{tabular}{|cl|lllll|lllll|}
			\hline
			\multicolumn{2}{|c|}{} &
			\multicolumn{5}{c|}{Regression Error} &
			\multicolumn{5}{c|}{Time Taken} \\ \hline
			\multicolumn{2}{|c|}{Threads} &
			\multicolumn{1}{l|}{8} &
			\multicolumn{1}{l|}{16} &
			\multicolumn{1}{l|}{32} &
			\multicolumn{1}{l|}{64} &
			128 &
			\multicolumn{1}{l|}{8} &
			\multicolumn{1}{l|}{16} &
			\multicolumn{1}{l|}{32} &
			\multicolumn{1}{l|}{64} &
			128 \\ \hline
			\multicolumn{1}{|c|}{\multirow{5}{*}{Scheme}} &
			FAA &
			\multicolumn{1}{l|}{22} &
			\multicolumn{1}{l|}{35} &
			\multicolumn{1}{l|}{30} &
			\multicolumn{1}{l|}{29} &
			40 &
			\multicolumn{1}{l|}{3961} &
			\multicolumn{1}{l|}{5525} &
			\multicolumn{1}{l|}{5759} &
			\multicolumn{1}{l|}{7842} &
			5100 \\ \cline{2-12} 
			\multicolumn{1}{|c|}{} &
			CAS &
			\multicolumn{1}{l|}{22} &
			\multicolumn{1}{l|}{33} &
			\multicolumn{1}{l|}{26} &
			\multicolumn{1}{l|}{25} &
			30 &
			\multicolumn{1}{l|}{9135} &
			\multicolumn{1}{l|}{8984} &
			\multicolumn{1}{l|}{12327} &
			\multicolumn{1}{l|}{16968} &
			3800 \\ \cline{2-12} 
			\multicolumn{1}{|c|}{} &
			Lock-Based &
			\multicolumn{1}{l|}{22} &
			\multicolumn{1}{l|}{33} &
			\multicolumn{1}{l|}{26} &
			\multicolumn{1}{l|}{25} &
			30 &
			\multicolumn{1}{l|}{1488} &
			\multicolumn{1}{l|}{1871} &
			\multicolumn{1}{l|}{2364} &
			\multicolumn{1}{l|}{3192} &
			4164 \\ \cline{2-12} 
			\multicolumn{1}{|c|}{} &
			Hogwild &
			\multicolumn{1}{l|}{1529} &
			\multicolumn{1}{l|}{1467} &
			\multicolumn{1}{l|}{1653} &
			\multicolumn{1}{l|}{1368} &
			1627 &
			\multicolumn{1}{l|}{3851} &
			\multicolumn{1}{l|}{5876} &
			\multicolumn{1}{l|}{6121} &
			\multicolumn{1}{l|}{7561} &
			5800 \\ \cline{2-12} 
			\multicolumn{1}{|c|}{} &
			LF &
			\multicolumn{1}{l|}{22} &
			\multicolumn{1}{l|}{33} &
			\multicolumn{1}{l|}{26} &
			\multicolumn{1}{l|}{25} &
			30 &
			\multicolumn{1}{l|}{514} &
			\multicolumn{1}{l|}{600} &
			\multicolumn{1}{l|}{183} &
			\multicolumn{1}{l|}{277} &
			207 \\ \hline
		\end{tabular}
	}
 \caption{Different Lock-free Regression Fitting}
	\label{table:collect}
\end{table}
\begin{table}[h]

\resizebox{\columnwidth}{!}{%
\begin{tabular}{|p{0.1\columnwidth}p{0.15\columnwidth}|c|c|c|c|c|c|}
\hline
\multicolumn{2}{|p{0.15\columnwidth}|}{Workloads}      & books    & fb        & osmc      & normal    & log normal & uniform\_sparse \\ \hline
\multicolumn{2}{|l|}{Number of Data} & 200M     & 200M      & 200M      & 200M      & 200M       & 200M                       \\ \hline
\multicolumn{1}{|l|}{\multirow{5}{0.14\columnwidth}{No. of Models or Nodes}} & LPA       & 330,571  & 1,572,094 & 3,139,126 & 1,603     & 2,244              & 71,161          \\ \cline{2-8} 
\multicolumn{1}{|l|}{}   & PGM       & 263,414  & 1,081,848 & 680,521   & 1,385     & 1,942               & 53,258          \\ \cline{2-8} 
\multicolumn{1}{|l|}{}   & \lfabt   & 27821943 & 67184623  & 15559343  & 113989225 & 39144243        & 18505423        \\ \cline{2-8} 
\multicolumn{1}{|l|}{}  & \cist      & 18963332 & 19592860  &    20094599       &      20033152     & 19563103   &     20666490                            \\ \hline

\end{tabular}%
}
\caption{Cardinality of Hierarchy}
\label{table:nodes}

\end{table}

%% file: Diagram/PGM.tex
\begin{figure}[]
\captionsetup[subfigure]{justification=centering}
\centering
  \includegraphics[width=0.99\linewidth]{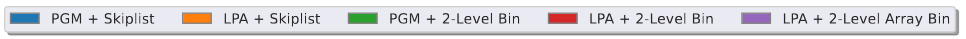}
  \subfloat[]{  \includegraphics[width=0.48\linewidth]{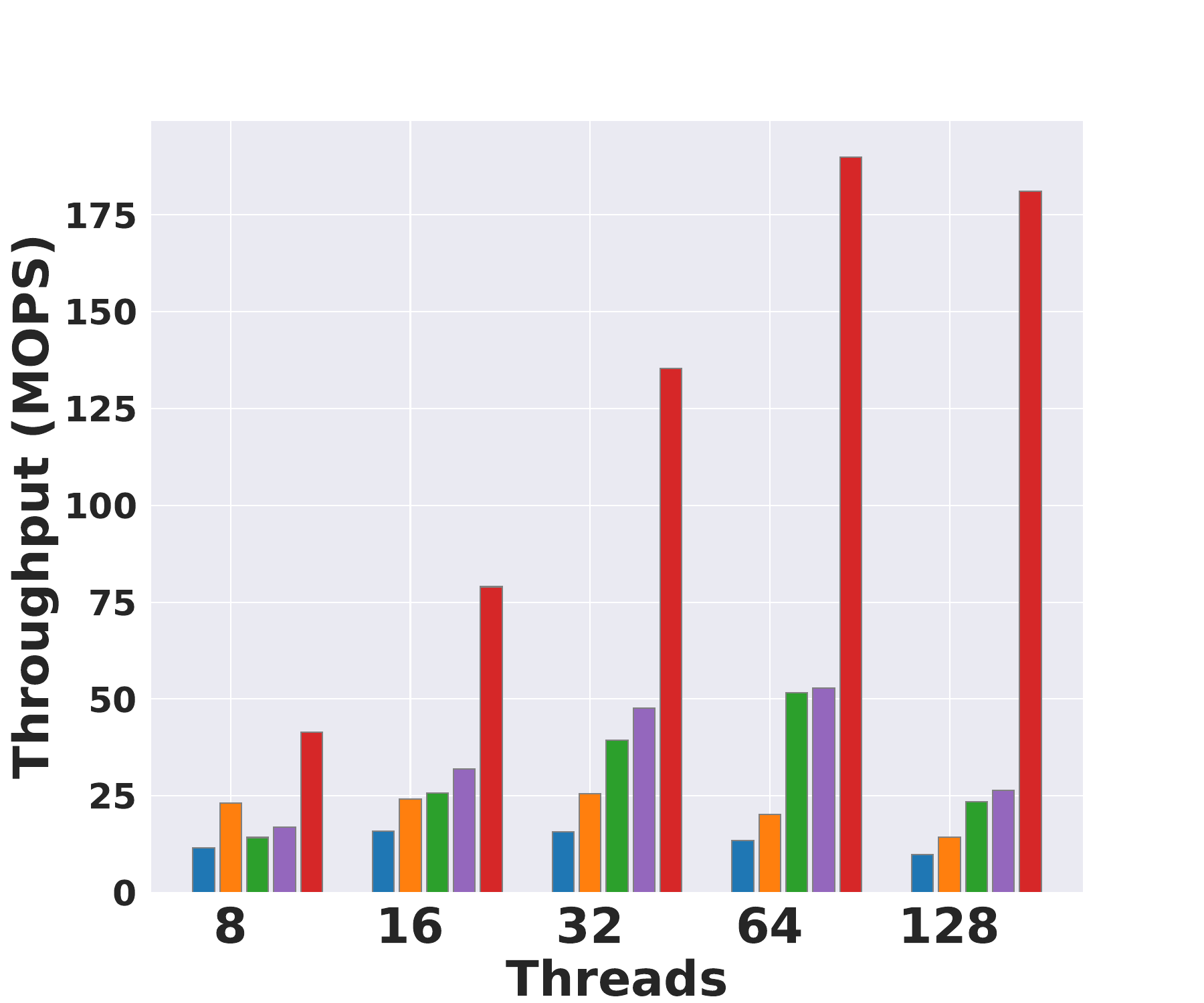}}
   \subfloat[]{ 
  \includegraphics[width=0.48\linewidth]{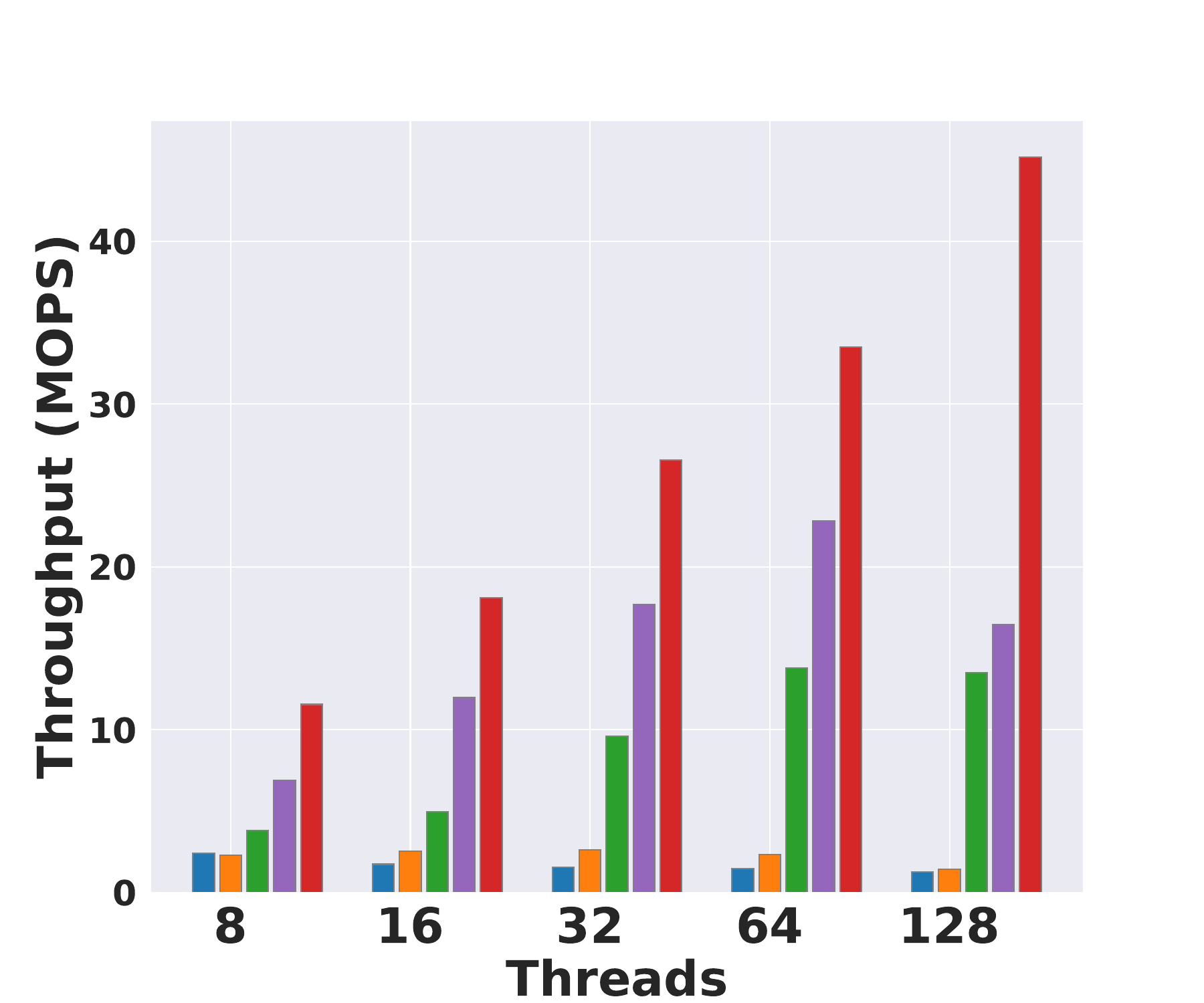}}
\caption{a) Throughput with read heavy workload with different combinations on uniform dataset b) Throughput with update heavy Workload with different combinations on uniform dataset  }
\label{fig:PGM}
\end{figure}

%% file: Sections/correct.tex
We start with observing some invariants maintained by the presented algorithm.

\begin{invar}\label{inv:unipath}
	There is a unique path from the root to a key.
\end{invar}  
As the traversal scheme uses models whose parameters are immutable, we can use induction to argue that any update to \kan does not invalidate Invariant \ref{inv:unipath}. As soon as the threshold of a \bin is reached, every update operation will either freeze its \knd{s} or will help at that. Once the \knd{s} are frozen, they become immutable, and any \insertADT or \remove operation necessarily engages in helping before reattempting its steps. In the helping phase, all \knd{s} of a \tbin are collected, and a model is trained. Only after a frozen \obin is replaced by a \tbin, or a frozen \tbin is replaced by a new \mnd, the search of a key inserted or deleted will return as defined. This ensures that Invariant \ref{inv:update} is maintained.
\begin{invar}\label{inv:update}
	\begin{enumerate}[noitemsep, topsep=0.2pt, leftmargin=*, nosep, nolistsep, label=(\alph*)]
		\item No key-value pair is lost due to the transformation of a bin in its lifecycle.
		\item A \search operation always returns the latest value paired with a key inserted in \kan. Once a key is deleted, a \search always returns \nul.
	\end{enumerate}
\end{invar}

The recursive call of \textsc{Scan} operation starting from the root ensures that it collects all the target key-value pairs. At the same time, a range search is oblivious to an ongoing transformation of a \bin object, which is only replaced using an atomic \cas. This ensures Invariant \ref{inv:rsearch}.
\begin{invar}\label{inv:rsearch}
	A range search returns every key-value pair defined by its query key and range size.
\end{invar}  

\myparagraph{Linearizability:} Having observed the maintenance of the above invariants, we discuss the linearization points (LPs). The LP of a successful \insertADT is at the atomic \cas to either insert a new \bin, a new \knd, or a new \vval object. Similarly, for a successful \remove operation, it is the \cas to introduce a \vval object with \nul value. For an unsuccessful \insertADT, it is at the atomic read step of the \vval object pointer that contains a matching value. For an unsuccessful \remove operation LP is at the invocation if the key was not present in the data structure or immediately after the concurrent \remove operation that would have removed the key. The LP of an unsuccessful \search operation is determined similar to an unsuccessful \remove operation. A successful search operation linearizes at the atomic read step of the \vval pointer containing the latest value associated with the query key. A \rns operation linearizes on atomically reading the global \texttt{Timestamp}. The linearizability of ADT operations is proved by ordering them in an arbitrary concurrent execution by their linearization points (LPs).

\myparagraph{Lock freedom:} We already observed that on a \cas failure, an \insertADT or \remove operation either retries or helps. The read operations do not either obstruct any operation or engage in helping. Evidently, there is no loop in their execution path. This establishes that at least one non-faulty thread will complete its operation in a finite number of steps in a concurrent execution, proving lock freedom.


%% file: Sections/Evaluation.tex
\myparagraph{Platform Configuration:} We conducted the experiments on a system with an AMD EPYC 7452 with 2 NUMA processing units packing 32 physical cores each thus with a total of 64 cores, with a minimum clock speed of 1.5 GHz and a maximum clock speed of 2.5 GHz. There are two logical threads for each core, and each has a private 32KB L1 data cache and L1 instruction cache. Every pair of cores shares a 512KB L2 cache and a 10MB L3 cache. The system has 256GB RAM and a 2TB hard disk. The machine runs Ubuntu 18.04.6 LTS. Our implementation is based on C++; the code was compiled using g++ 11.1.0 with -std=c++17 and linked the pthread and atomic libraries. 

\myparagraph{Experiments Setup:} In each case, we take a dataset of 200M keys. We prefill the data structure with 10M keys randomly selected from the dataset. Then we take a random permutation of the dataset to perform ADT operations \insertADT, \remove, and \search. Each experiment is performed by warming up the system, and then steady results are taken.

\input{Diagram/results}

\myparagraph{Algorithms:} We compared \kan with state-of-the-art comparison-based and learned indexes: (a) Lock-free $(a,b)$-tree (\lfabt) \cite{TrevorBrown:PhDThesis}, (b) Elimination $(a,b)$-tree (\eabt) \cite{srivastava2022elimination}, (c) lock-free interpolation search tree (\cist) \cite{brown2020non}, and (d) \fdex \cite{li2021finedex}. We did not include XIndex \cite{tang2020xindex}, as \fdex already reported 1.3x speed up over them across the workloads and data distributions.

\myparagraph{Memory Management} has a significant overhead on the performance of lock-free data structures, which lock-based ones -- \eabt and \fdex -- escape. For a fair comparison, we include two variants of \kan, one with memory management -- \mrkan, and another without that. We used the epoch-based memory management scheme DEBRA \cite{brown2015reclaiming} as used by \lfabt and \cist.

\myparagraph{Datasets and Distributions} We used the datasets and distributions used in \cite{kipf2019sosd} and \cite{srivastava2022elimination} to evaluate \kan vis-a-vis its competitors. Each dataset includes 200 million 64-bit unsigned integers, out of which 10 million are initially used to populate the data structure; the operations after initialization use the complete dataset. More specifically, we used (1) Amazon \cite{Amazon_Dataset},  (2) Facebook~\cite{van2019efficiently}, (3) Wikipedia~\cite{Wiki}, and (4) OSM~\cite{pandey2018good}, in addition to synthetic datasets generated using uniform and lognormal distributions. Amazon is the book sale popularity data; Facebook represents the unsampled version of the Facebook user ID dataset; Wikipedia is the article edit timestamps; OSM is uniformly sampled OpenStreetMap locations represented as Google S2 CellIds. Finally, we also used Yahoo Cloud Service Benchmark \cite{cooper2010benchmarking} (YCSB) workloads with uniform as well as Zipfian distribution. Notice that, such a selection of datasets ensures \textit{critical testing of the proposed scheme under real-life and skewed data} settings.

\subsection{Read-heavy workload} 
Having initially populated the data structure, concurrent threads perform ADT operations for 10 seconds out of which 95\% are \search, 3\% are \insertADT, and 2\% are \remove. We record the throughput in million operations per second vs. the number of threads for each of the algorithms. 
 

We examine the scalability with number of threads: 8, 16, 32, 64 and 128, on uniformly distributed data as plotted in Figure \ref{fig:results} (a). Across the methods, throughput increases as we increase the number of threads from 8 to 64. The impact of hyperthreading (128 threads) is observed on each of them. Throughout \kan outperforms its competitors by handsome margins, in particular, it offers approximately two orders of magnitude better performance over the compared lock-free schemes -- \lfabt and \cist, even after discounting for memory management overhead. Examining the methods without memory management, \kan outperforms \eabt by 10x for 8 and 16 threads and up to 4x for 64 threads. Owing to its lock-free progress, \kan significantly outperforms \fdex as contention increases with the number of threads.

\kan maintains an excellent lead over its competitors across the datasets as we see in Figure \ref{fig:results} (b). In this set of plotted results we used 64 threads, however, the comparative performance is unchanged with other contention levels as well. It offers an average of around 1.25x higher throughput in comparison to its nearest competitor \fdex.

To understand these results, we note that the comparison-based structures use the traditional search technique to traverse a hierarchy of height $O(log(n))$, by contrast, \fdex and \kan, not only use $O(1)$ arithmetic operations for a model inference at each level but also traverse a much shallower hierarchy. As \fdex uses fine-grained locks on bins for every query, \kan outperforms it for its lock-free scheme for every query. Though \cist employs interpolation search in internal 'fat' nodes, it still performs poorly because the cost of rebalancing is too high even with a small number of updates in the workload. 

\subsection{Update-heavy Workload} 
This set of workloads is constructed in the same way as the read-heavy case except that a 10 seconds run of operations comprises of 50\% \insertADT, 20\% \remove and 30\% \search. The performance measurement method remains as it was earlier.

Performance scalability with threads on uniform dataset is plotted in Figure \ref{fig:results} (c). We note that there is no degradation in performance even after hyperthreading kicks in. Irrespective of the number of threads, \kan significantly outperforms its competitors for identical memory management.
 
Figure \ref{fig:results} (d) demonstrate a comparative performance of \kan with its competitors using 64 threads. Compared to read-heavy workload, the conventional data structures \cist and \lfabt catch up with \kan, which still offers a throughput as good as twice that of \lfabt and four times that of \cist across the datasets. Furthermore, without memory management, it outperforms \fdex with a decent margin in all types of data distributions.

We can understand this performance as the following. Conventional data structure performed well despite traversing down the tree because the leaf node size, where the dataset updates are ingested, is much smaller compared to \kan's bin. The bin size in \kan should be big enough to train the model over it to avoid the effort of retraining repeatedly. Even in this case, \kan  owes its lead by an average 1.17x higher throughput over \fdex to its lock-free design. 

\subsection{YCSB Workload}
We benchmarked \kan and its competitors on the standard YCSB workloads -- which are of three levels A: update-heavy, B: read-heavy, and C: read-only -- with uniform and Zipfian distributions. The results are plotted in Figures \ref{fig:results} (e) and (f). 

With uniform distribution, it surpassed \fdex in update-heavy workload by 1.13x, whereas \eabt outperformed \fdex by 1.08x. In read-heavy workload, \kan outperformed \eabt by 1.34x. In read-only workload, \fdex performed as good as \kan. \kan outperformed \eabt by 1.37x in read-only workload.

Similarly, \kan outperformed its competitors with Zipfian distribution.  The performance of \fdex dropped by 4.8x in Zipfian distribution compared to the uniform distribution due to its locks. Similarly, in read-heavy workload, \eabt outperformed \fdex by 1.19x, whereas \kan outperformed \eabt by 1.14x. In read-only workload, \kan outperformed \fdex by 1.07x. 

It is interesting to observe that with better concurrency management, such as with an elimination scheme, even a comparison-based index, such as \eabt, can significantly outperform an efficient learned index, such as \fdex, over an adverse workload, which is the case with YCSB zipfian update-heavy workload. In any case, \kan offers better performance compared to competitors across the workload for identical memory reclamation overheads. Clearly, \kan brings together the best of both worlds, i.e. lock-free synchronization and learned queries.

\subsection{LLC Misses}
We also compared the LLC misses of the data structures on uniform distribution for both read-heavy and update-heavy workloads in Figure \ref{fig:results} (g) and (h), respectively. LLC misses of different data structures corroborate their respective throughput. 
The big fat nodes of \kan make it more cache efficient than its conventional competitors, where the depth is high to reach the leaf node to perform the operations. \cist, which again uses the fat nodes at the internal level with less depth of leaf node from the root node, still receives the highest number of LLC misses due to its complicated rebalancing mechanism. On the other hand, \kan gets an advantage here as it does not need to perform any rebalancing. Overall, \kan and \fdex demonstrate the least cache misses as compared to others due to their design.

\subsection{Throughput with Skewed Contention}
\input{Diagram/Hotspot}
To capture the effect of \textit{skewness in query-key distribution}, we  also checked the performance of the considered methods with different hotspot ratios, which indicates how the range of the available dataset in a data structure relates to the range of the query keys. Essentially, a low hotspot ratio indicates high contention among operations as taking place at a narrow portion of the data structure. The results are plotted in Figures \ref{fig:Hotspot} (a) and (b). The conventional comparison-based structures demonstrate a lesser effect of skewness, whereas  \fdex performs very poorly even with a hotspot ratio of 0.8, because of its locking technique on nodes even for read operations. The lack of a balancing mechanism in \kan results in the observation of comparative performance thrashing here; however, it can still outperform its competitors in most settings.

\subsection{Throughput with Range Search}
\begin{figure}[]
\captionsetup[subfigure]{justification=centering}
\centering
  \includegraphics[width=0.8\linewidth]{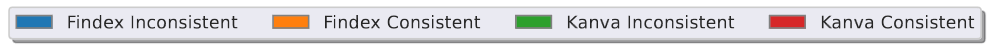}
  \subfloat[]{  \includegraphics[width=0.48\linewidth]{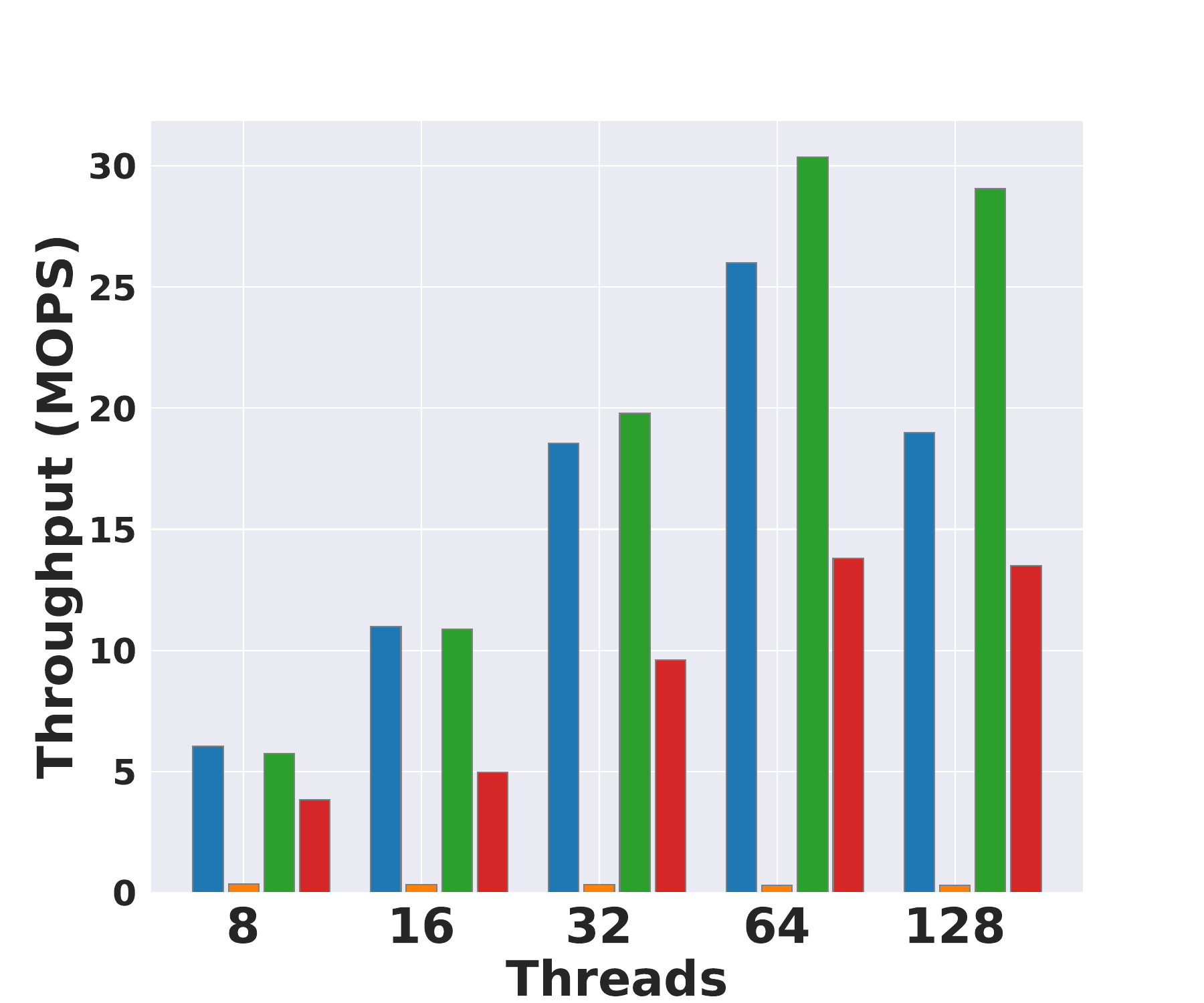}}
   \subfloat[]{ 
  \includegraphics[width=0.48\linewidth]{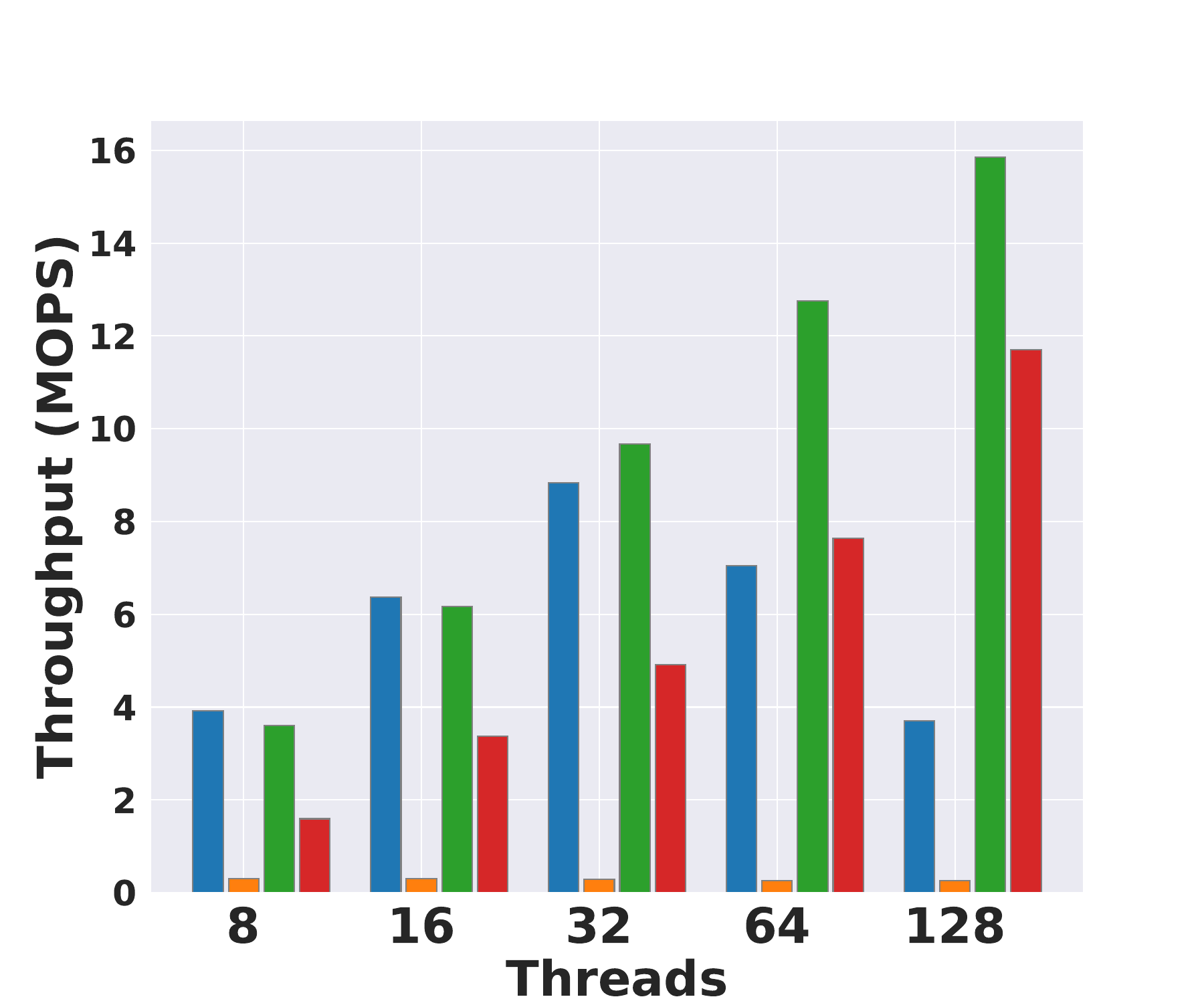}}
\caption{a) Throughput of read-heavy workload with Range Search b) Throughput of update-heavy workload with Range Search }
\label{fig:range_search}
\end{figure}

For consistent and inconsistent range queries, we compared the throughput of \kan with \fdex in Figure \ref{fig:range_search}(a) for read-heavy workload and \ref{fig:range_search}(b) for the update-heavy workload. In \fdex, fine-grained locks are used for inconsistent range search while coarse-grained locks are used for consistent range search. \fdex consistent range queries do not scales due to their coarse-grained locks, and \kan consistent and inconsistent range queries outperform it by a magnitude of degree 2. \kan also outperforms \fdex in inconsistent range query with a significant margin.

%% file: Diagram/results.tex
\begin{figure*}[ht]
\captionsetup[subfigure]{justification=centering}
\centering
  \includegraphics[width = 0.80\linewidth]{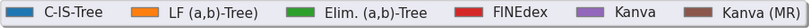}
  \subfloat[]{  \includegraphics[width=0.24\linewidth]{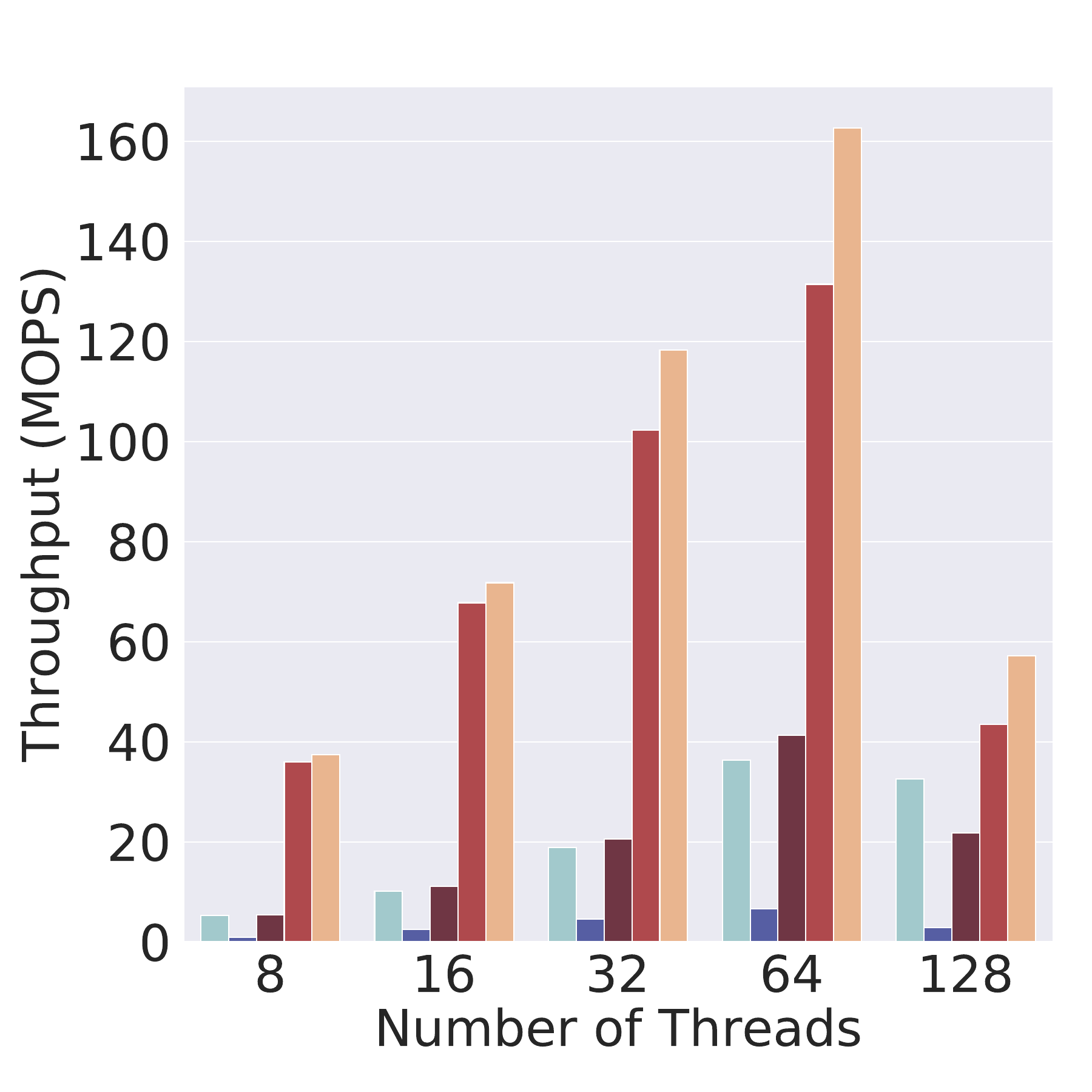}}
   \subfloat[]{ 
  \includegraphics[width=0.24\linewidth]{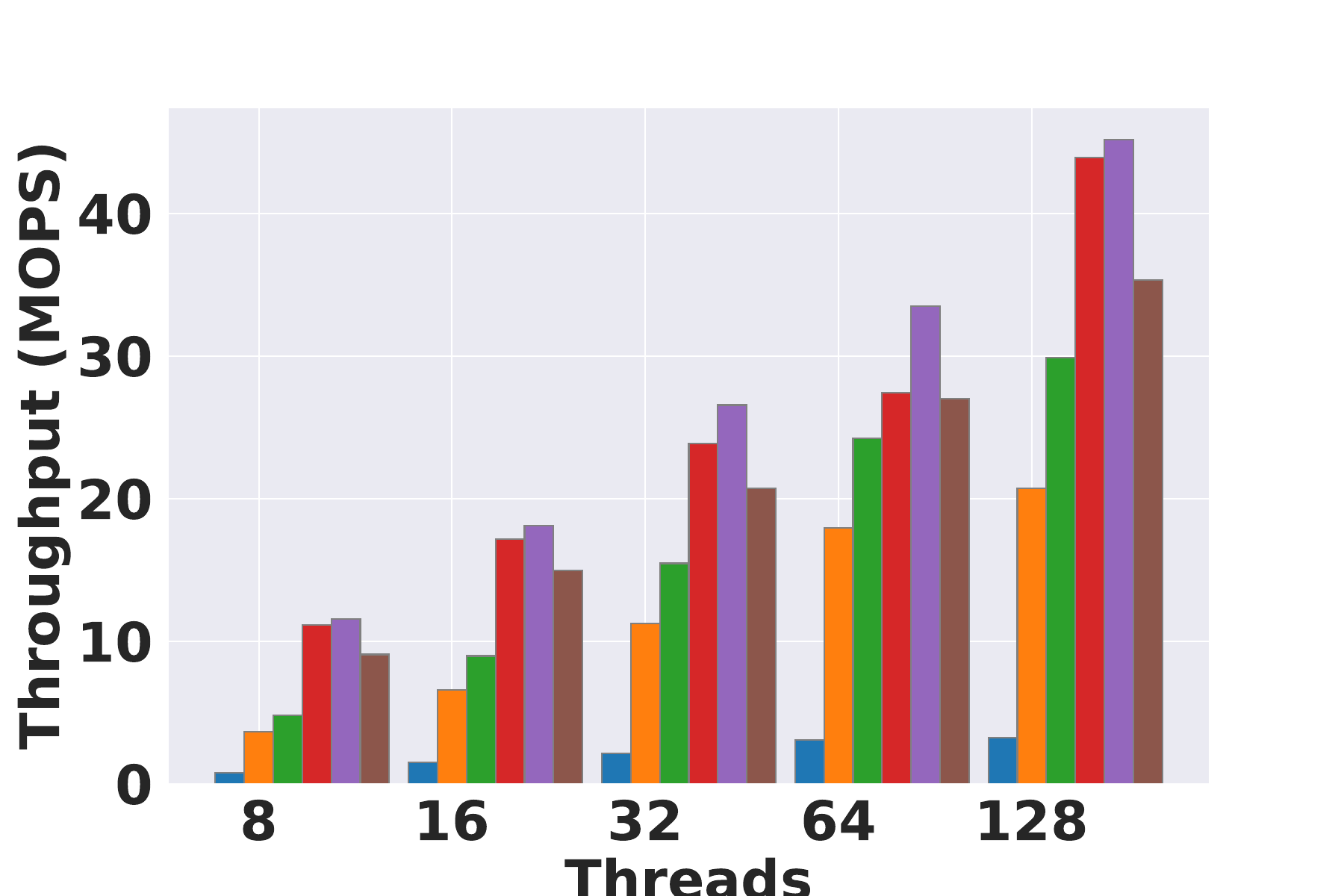}}
   \subfloat[]{ 
  \includegraphics[width=0.24\linewidth]{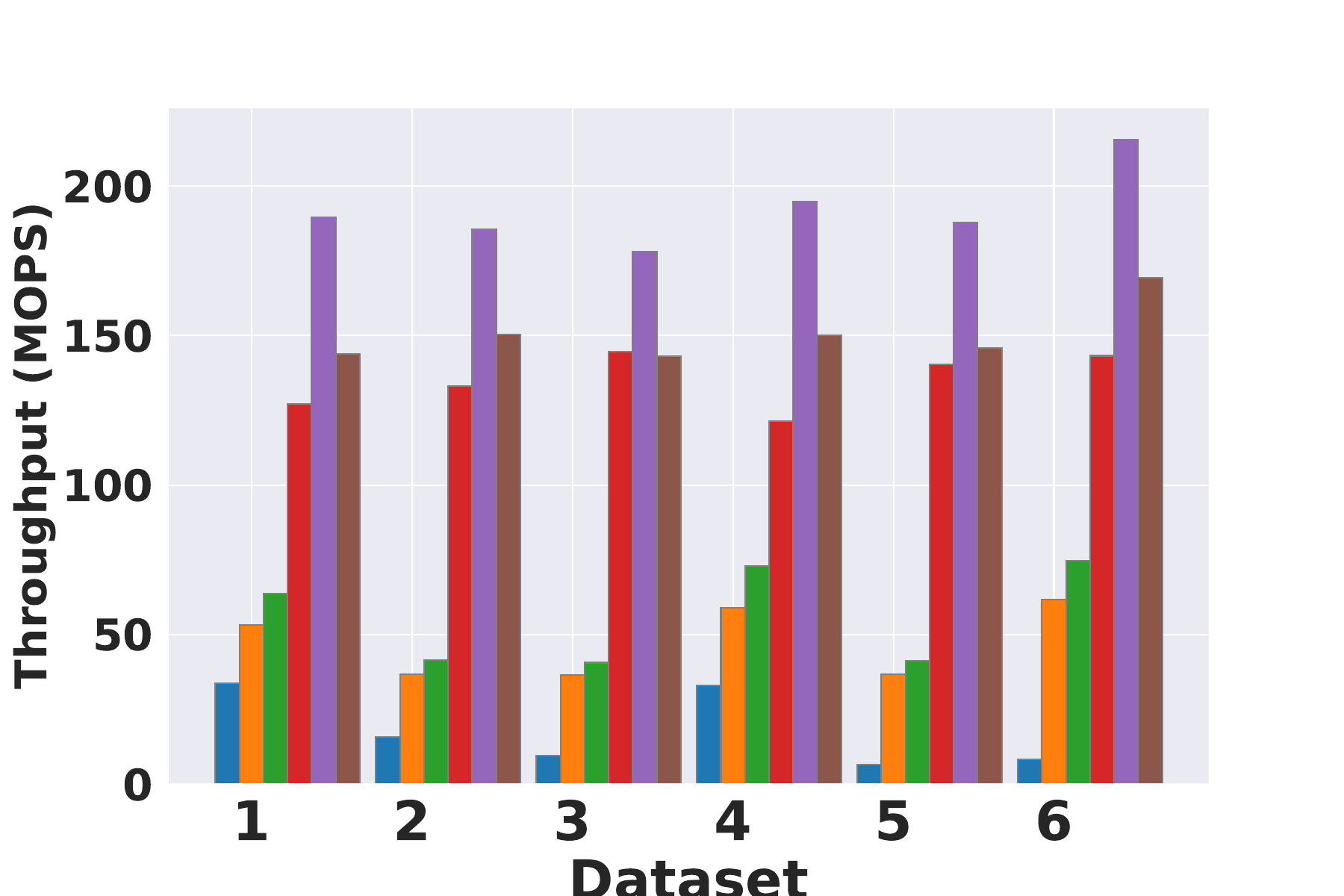}}
    \subfloat[]{ 
  \includegraphics[width=0.24\linewidth]{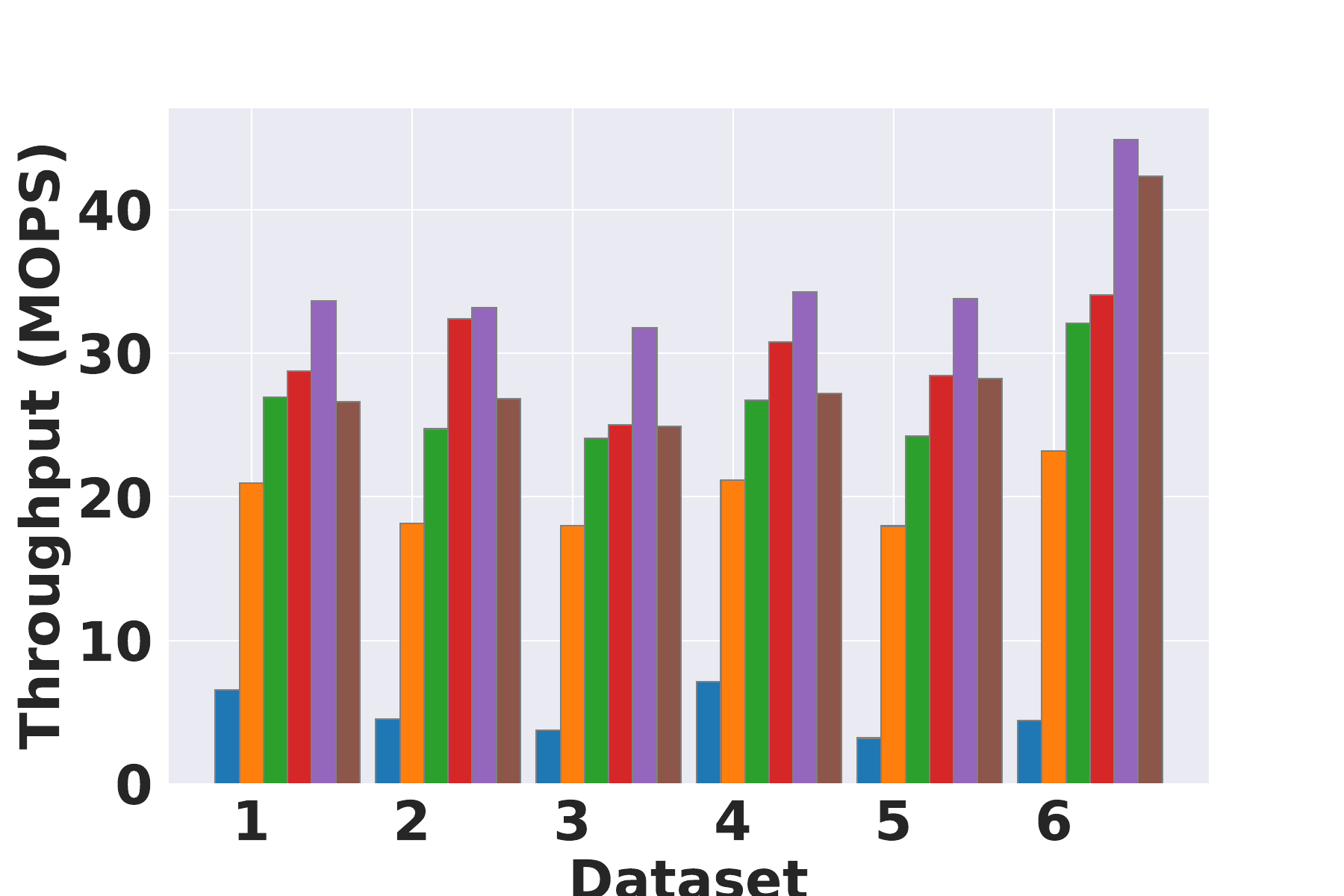}}
\vspace{1em}
   \subfloat[]{ 
  \includegraphics[width=0.24\linewidth]{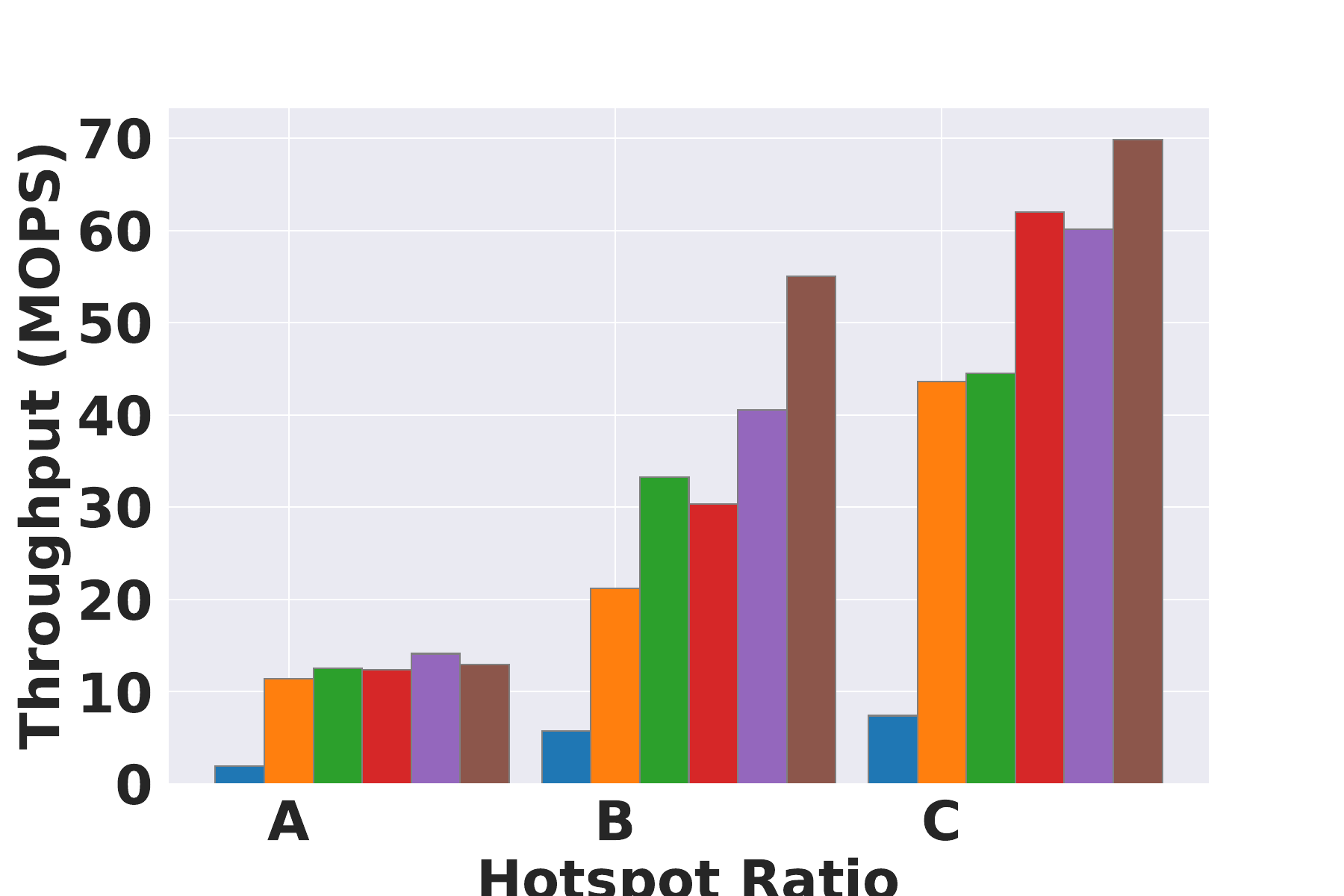}}
   \subfloat[]{ 
  \includegraphics[width=0.24\linewidth]{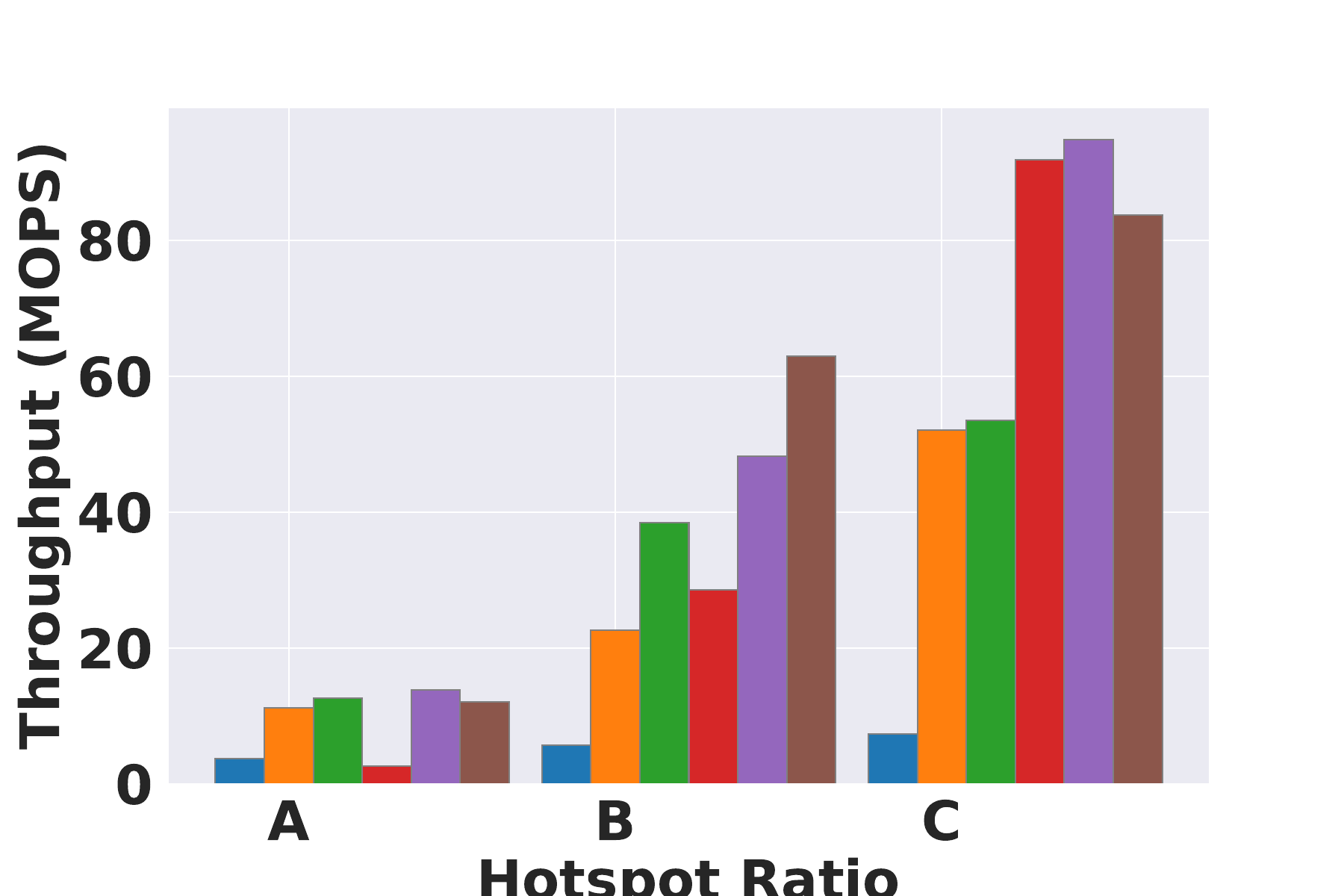}}
    \subfloat[]{ 
  \includegraphics[width=0.24\linewidth]{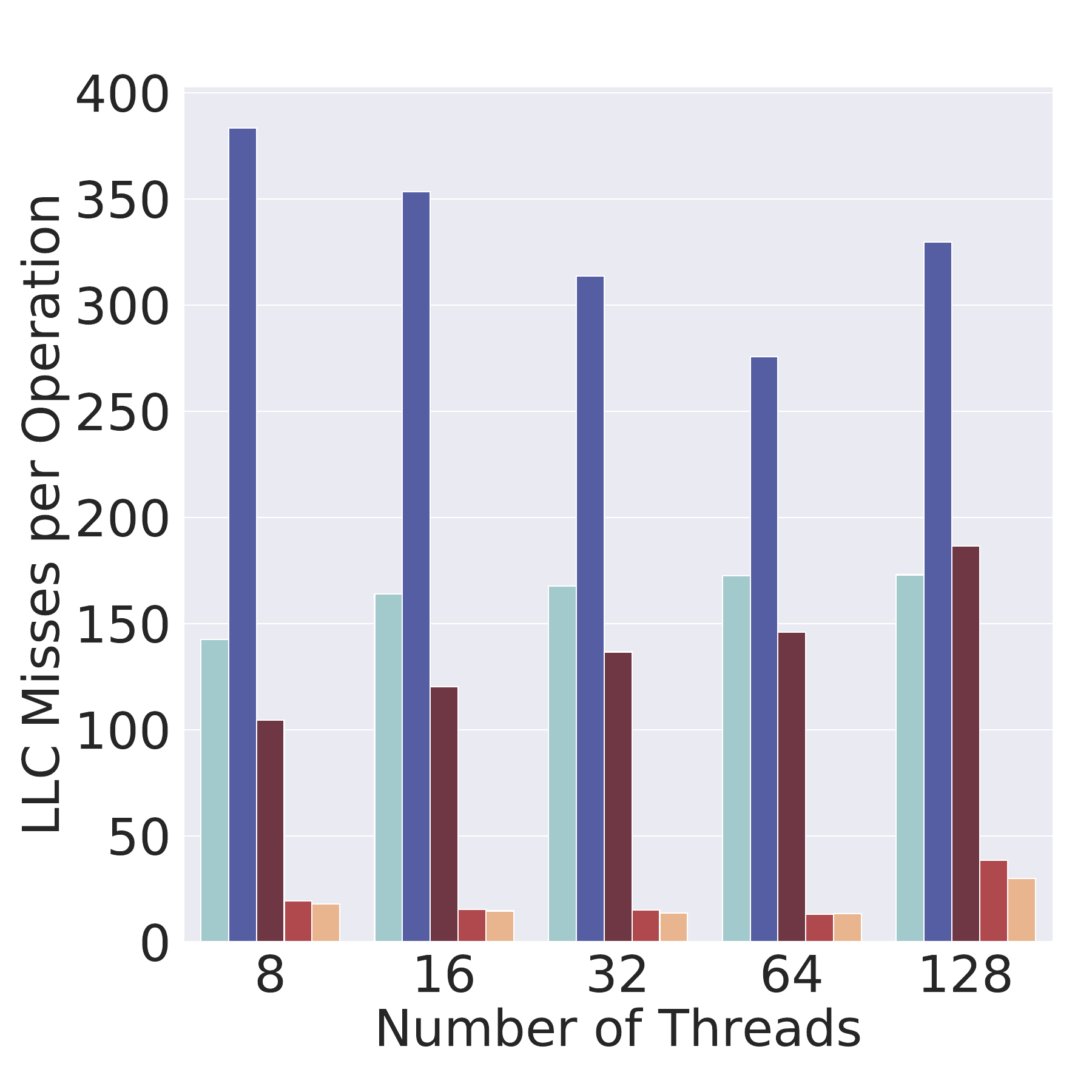}}
   \subfloat[]{ 
  \includegraphics[width=0.24\linewidth]{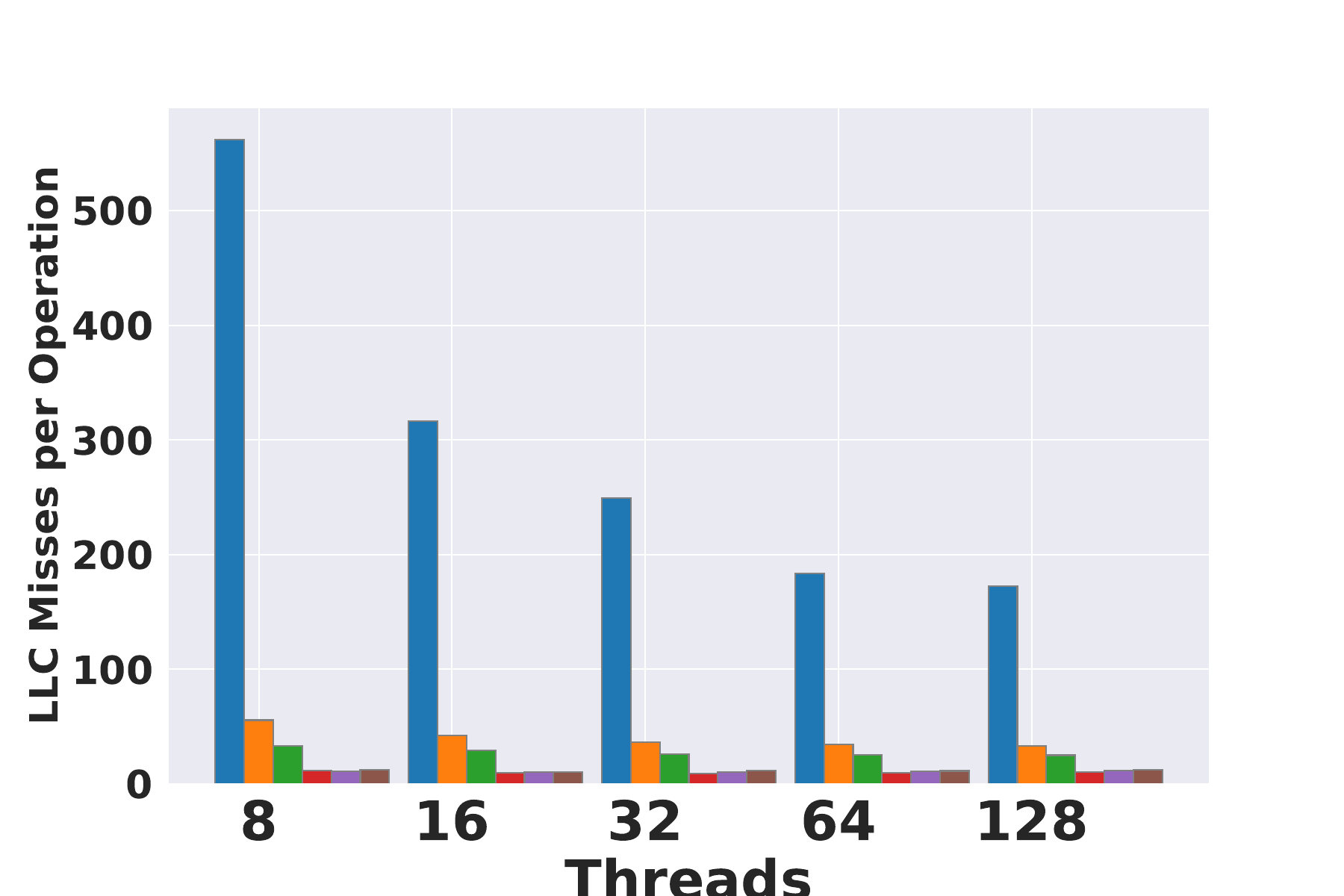}}
\caption{ The performance of \textbf{Kanva} when compared to different datastrucutre. (a) read heavy workload with uniform dataset (b) update heavy Workload with uniform dataset (c) read Heavy workload with different dataset (1: uniform, 2: normal, 3: Amazon, 4: Facebook, 5: OSM(Open Street Map), 6: Wikipedia) (d) update heavy workload with different dataset (e) YCSB workload with uniform distribution (f) YCSB workload with ziphian distribution (g) last-level cache misses (lower is better) on read heavy workload, (h) last level cache misses on update heavy workload}
\label{fig:results}
\end{figure*}

%% file: Diagram/Hotspot.tex
\begin{figure}[]
\captionsetup[subfigure]{justification=centering}
\centering
  \includegraphics[width=0.8\linewidth]{Diagram/Screenshot_from_2023-08-05_17-31-21.png}
  \subfloat[]{  \includegraphics[width=0.48\linewidth]{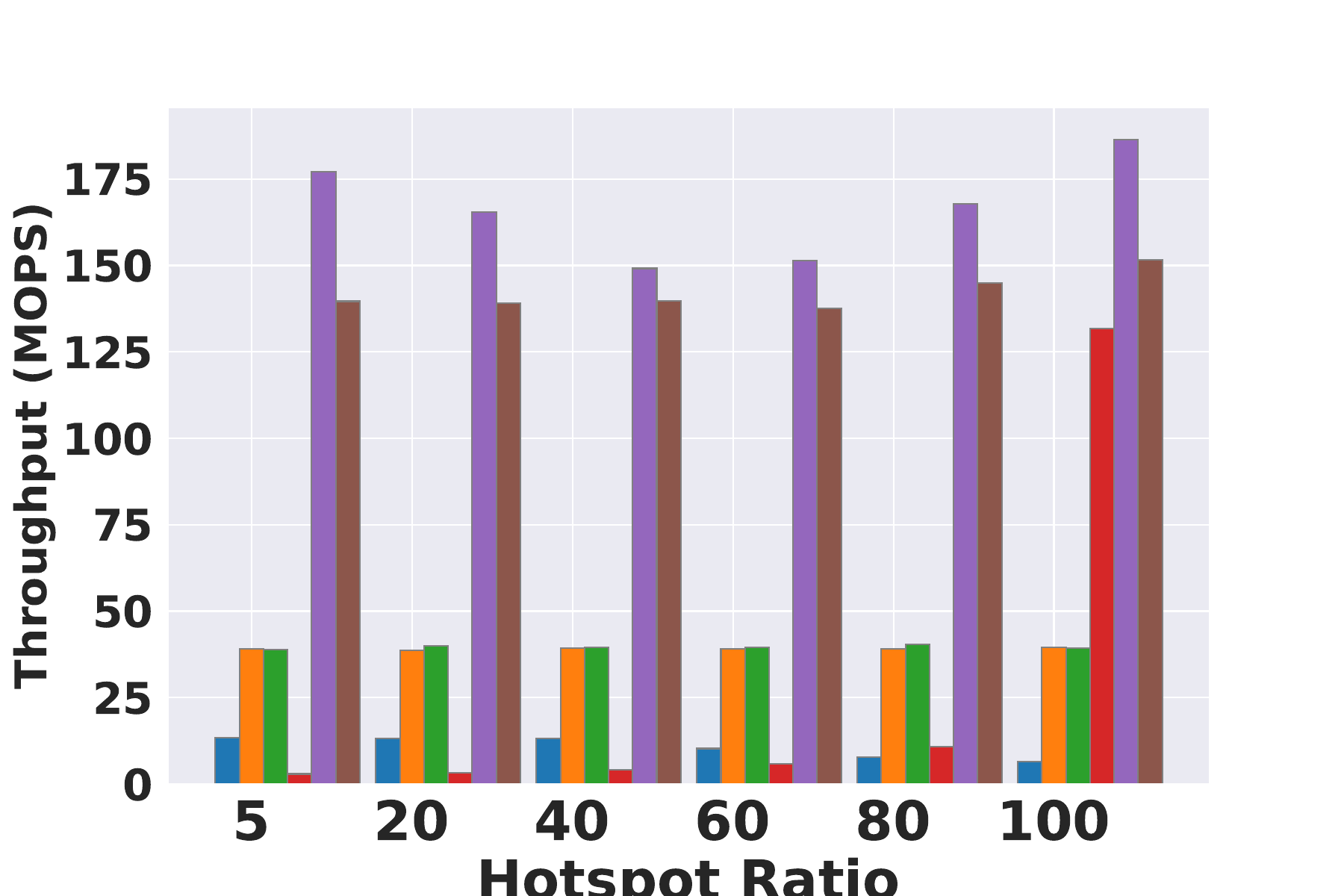}}
   \subfloat[]{ 
  \includegraphics[width=0.48\linewidth]{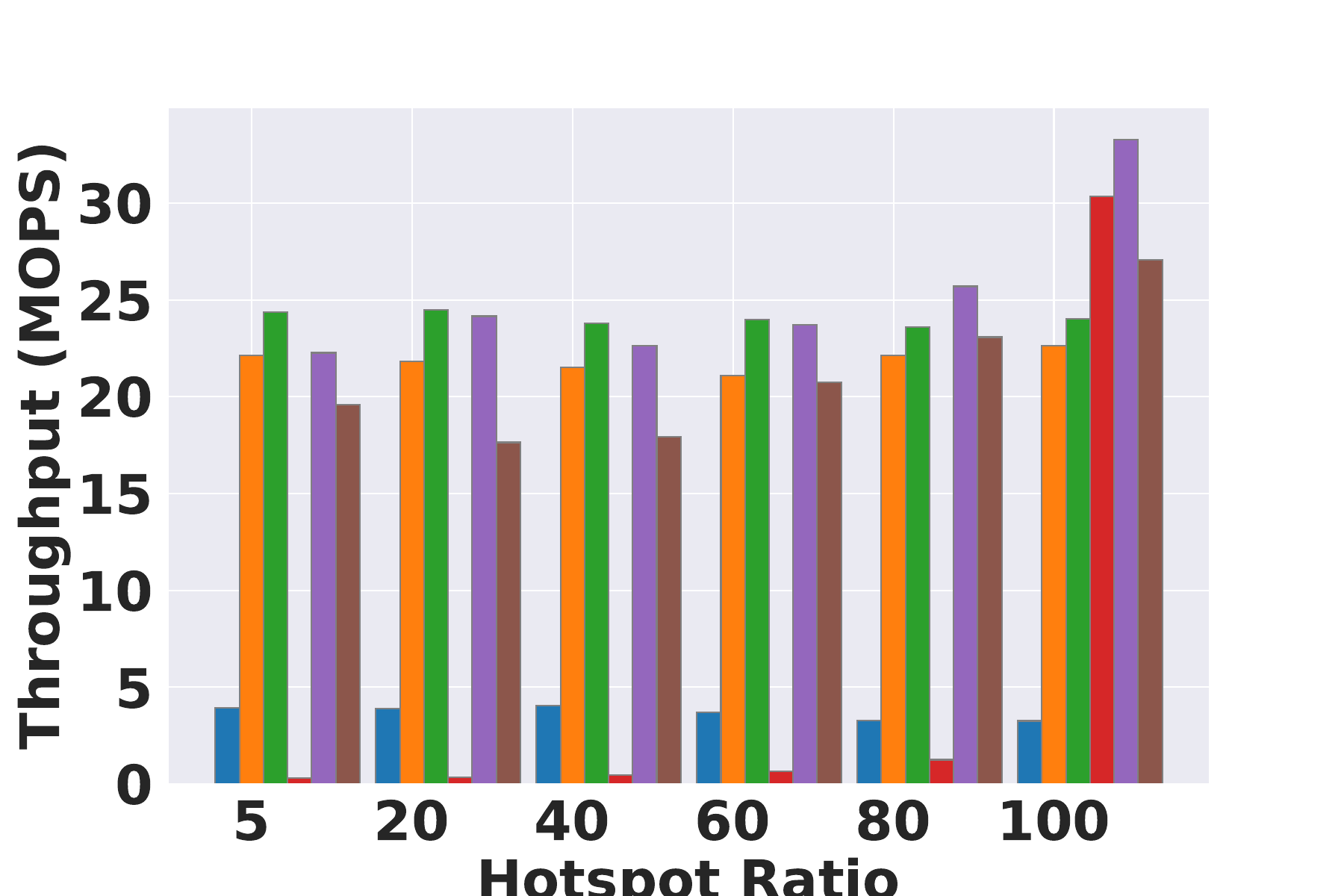}}
\caption{a) Throughput with read heavy workload with different hostpot Ratio b) Throughput with update heavy workload with different hostpot Ratio }
\label{fig:Hotspot}
\end{figure}

%% file: Sections/BackgroundandMotivation.tex
\textbf{B+Tree and variants} are classical indexes where each $e_i\in\mathcal{E}$ is based on comparison operations with $B$ data points, stored in ``fat'' nodes. Some of the lock-free designs are lock-free B$^+$-tree \cite{braginsky+:SPAA:2012}, Open BW-Tree~\cite{wang2018building}, and BZ-Tree \cite{arulraj2018bztree}. 

\textbf{$(a,b)$-Tree} derives from B$^+$-trees by relaxing the size of the fat nodes to vary between $a$ and $b$, whose lock-free implementation was introduced by \cite{TrevorBrown:PhDThesis}. Recently, Elimination (a,b) tree \cite{srivastava2022elimination}, a lock-based variant of ~\cite{TrevorBrown:PhDThesis}, which also uses elimination \cite{hendler2004scalable} of concurrent \insertADT and \remove operations with the same key, demonstrably outperformed several other classical concurrent indexes. 


Lock-free interpolation search tree \textbf{C-IST}~\cite{brown2020non} uses the interpolation search \cite{mehlhorn1993dynamic} to choose the child node for indexing. They are our nearest classical comparison-based index counterpart. Given a dataset of size $n$, the root node is of degree $\sqrt{n}$ covering the entire dataset and dividing it into $\sqrt{n}$ buckets. The child node of a bucket covering $\sqrt{n}$ keys will have $\sqrt[4]{n}$ degree. This rule-based division continues for $O(log~log~n)$ levels, ensuring expected amortized $O(log~log~n + p)$ steps for data structure operations on smooth key distributions, where $n$ is the dataset size at the invocation, and $p$ is the maximum number of threads. The data exists at leaf nodes. Interpolation search makes the indexing faster for favourable key distributions; however, rebalancing is costly in C-IST. The fat nodes are array-based and are reallocated and replaced using double-compare-single-swap (DCSS) for every update; however, the operations display good cache behaviour. By contrast, we never reallocate \mnd{s} and use only system native single-word \cas primitives.

\cite{wang2018building} and ~\cite{arulraj2018bztree} presented experimental results for range search queries. However, their implementations\textit{ are not \lble}. Others \cite{braginsky+:SPAA:2012}, \cite{TrevorBrown:PhDThesis}, \cite{srivastava2022elimination}, \cite{brown2020non} do not have any implementation of range search, though \cite{srivastava2022elimination} and \cite{brown2020non} mention that it can be implemented following \cite{brown2022pathcas}.

\ignore{\subsection{Learned Indexes}
\textbf{Recursive model index} (RMI) proposed by \cite{kraska2018case} uses a hierarchy $e_i\in\mathcal{E}$ of multi-layer perceptrons. A model predicts the model at the next level of hierarchy, thus providing a path to an approximate location of the query key in the sorted array of datasets at the leaf level. Models are chosen based on the key distribution, which also sets up a trade-off between the complexity of the model hierarchy and the prediction accuracy. The models are trained level by level, starting from the lowest one. RMI does not allow updates in the dataset. While the model hierarchy training in RMI requires accessing the dataset multiple times, \textbf{PGM-index}~\cite{ferragina2020pgm}, \textbf{FITing-tree}~\cite{galakatos2019fiting} and \textbf{Radix-Spline}~\cite{kipf2020radixspline} fit models in a single pass over the data with a fixed error bound, which reduced the index building cost. These schemes, in static cases, offer performance competitive with RMI.

ALEX~\cite{Alex} and dynamic PGM-Index~\cite{ferragina2020pgm} allow dynamic updates. The updates ingested in gapped arrays in ALEX or LSM-managed updates in PGM-Index naturally required retraining the models. In general, these sequential indexes implement range search as well. 
}
\textbf{XIndex} ~\cite{tang2020xindex} was the first learned index to support concurrent updates. Similar to classical B$^+$-trees, the dataset is stored in leaf nodes. The index uses the RMI \cite{kraska2018case}. Each leaf node contains an associated model for search. In addition, they contain buffers for ingesting updates. 
 A query in a leaf node has to explore both the data and the buffer whereof searching in the buffer uses a comparison-based scheme. It uses background threads to merge the data and the buffer under barriers to maintain consistency, which blocks the update operation in the buffer. 

\textbf{\fdex}~\cite{li2021finedex}, which \kan builds on, addressed the drawbacks of XIndex. Firstly, in the leaf nodes, they used "bins" as in \kan, that enabled both cache-optimized read and lock-based concurrent updates, scrapping the need to explore the data array and buffer separately. 
They used a heuristic approach~\cite{xie2014maximum} to train the linear models with fixed error bound. 
\fdex outperformed XIndex across the workloads and data distributions. As mentioned in Section \ref{sec:intro}, both \cite{tang2020xindex} and \cite{li2021finedex} support only non-\lble range search.

%% file: Sections/conclude.tex
This paper presented the first \textit{lock-free learned search data structure}, which we named \kan. Our algorithm is provably linearizable.  Experimentally, \kan outperforms the existing lock-free and lock-based concurrent search methods by a good margin. In linearizable concurrent data structures, the closest competitor to the presented algorithm is Elimination-$(a,b)$-tree, which recently demonstrated superior performance to all existing concurrent search algorithms. With this relative comparison, the proposed algorithm can be placed as outperforming its existing competitors. 

Several instances exist in the literature on concurrent data structures where a carefully optimized lock-based scheme significantly outperforms the lock-free approach of translating a sequential data structure to a concurrent setting. A good example is the comparison between \eabt and \lfabt. Thus, it is refreshing to see a lock-free data structure outperforming every other existing concurrent data structure of the same class. 

Our experiments provided several insights. For example, a lock-based algorithm puts a fat node-wide lock to update a cell of the node array. Under this lock, it can shift elements without reallocating memory unless absolutely needed, for example, to maintain the node size threshold. By contrast, a lock-free scheme finds it challenging to synchronize without reallocating the node most of the time and atomically replacing it using a DCSS primitive or CAS primitive, with the former having some advantage in a few cases. There is a clear trade-off between the node size and synchronization options with the growing contention. Though \kan and \fdex have a similar index structure, \kan outperformed \fdex in both low and high contention cases. We obtained this sweet spot of node size by random search.


It is important to highlight that our experiments with skewed query-key distributions (small hotspot ratio) also revealed that an efficient synchronization method, such as one in \eabt, can enable even a comparison-based index to outperform a learned index, such as \fdex, as soon as the overhead of contention dominates the efficiency gained in query complexity. Nevertheless, our proposed method \kan sailed through this adverse examination as it combined efficient learned query and better progress guarantees of lock-free synchronization. This finding is worth noticing for future research in lock-free data structures, or for that matter, the same in concurrent learned indexes. 

Our technique is the first to allow lock-free range searches over learned data structures. We underscore that \kan outperformed \fdex by a wide margin, which demonstrated to have outperformed its competitors. 